\documentclass[notitlepage,aps,prd,twocolumn,superscriptaddress,showpacs]{revtex4-2}

\usepackage{natbib}
\usepackage[english]{babel}
\usepackage{graphicx}

\usepackage{amsmath, amssymb}
\usepackage{mathtools}
\usepackage[pdfborderstyle={/S/U/W 1}]{hyperref}
\usepackage{bbold}
\usepackage{empheq}
\usepackage{booktabs}
\usepackage[most]{tcolorbox}
\usepackage{siunitx}
\usepackage{tensor}
\usepackage{latexsym}
\usepackage{placeins}
\usepackage{csquotes}
\usepackage[acronym]{glossaries}
\glsdisablehyper
\usepackage{xfrac}
\usepackage{multirow}
\usepackage{titlesec}
\usepackage{csquotes}
\usepackage[normalem]{ulem}
\usepackage{colortbl}
\usepackage[caption=false,justification=centerlast]{subfig}
\usepackage{todonotes}
\usepackage{hhline}
\usepackage{physics}

\DeclareMathSizes{10}{10}{8}{6}
\DeclareSIUnit \persqrthz { /\sqrt{\si{\hertz}} }
\renewcommand{\arraystretch}{1.5}
\allowdisplaybreaks

\titleformat{\section}[runin]{\normalfont\bfseries}{\thesection.}{0.5em}{}[ -]
\titlespacing{\section}{0pc}{5mm}{2mm}

\definecolor{Gray}{gray}{0.9}

\everymath{\displaystyle}

\usepackage[most]{tcolorbox}

\newtcbox{\mymath}[1][]{%
    nobeforeafter, math upper, tcbox raise base,
    enhanced, colframe=black,
    colback=white, boxrule=1pt,
    #1}
    
\makeglossaries
    
\newacronym{e2e}{E2E}{End-To-End}
\newacronym{inrep}{INREP}{Initial Noise REduction Pipeline}
\newacronym{tdi}{TDI}{Time Delay Interferometry}
\newacronym{ttl}{TTL}{Tilt-To-Length}
\newacronym{dfacs}{DFACS}{Drag-Free and Attitude Control System}
\newacronym{ldc}{LDC}{LISA Data Challenge}
\newacronym{ldpg}{LDPG}{LISA Data Processing Group}
\newacronym{lsg}{LSG}{LISA Science Group}
\newacronym{lig}{LIG}{LISA Instrument Group}
\newacronym{simwg}{LDPG-SimWG}{Simulation Working Group}
\newacronym{aei}{AEI}{Albert Einstein Institute}
\newacronym{pssl}{PSSL}{Precision Space Systems Laboratory}
\newacronym{apc}{APC}{AstroParticule et Cosmologie}
\newacronym{syrte}{SYRTE}{Systèmes de R\'ef\'erence Temps-Espace}
\newacronym{l2it}{L2IT}{Laboratoire des deux Infinis de Toulouse}
\newacronym{irap}{IRAP}{Institut de Recherche en Astrophysique et Plan\'etologie}
\newacronym{lisa}{LISA}{Laser Interferometer Space Antenna}
\newacronym{emri}{EMRI}{Extreme Mass Ratio Inspiral}
\newacronym{ifo}{IFO}{Interferometry System}
\newacronym{grs}{GRS}{Gravitational Reference Sensor}
\newacronym{tmdws}{TM-DWS}{Test-Mass Differential Wavefront Sensing}
\newacronym{ldws}{LDWS}{Long-arm Differential Wavefront Sensing}
\newacronym[	plural={MOSAs},
		        first={Moving Optical Sub-Assembly},
		        firstplural={Moving Optical Sub-Assemblies}
            ]{mosa}{MOSA}{Moving Optical Sub-Assembly}
\newacronym{siso}{SISO}{Single-Input Single-Output}
\newacronym{mimo}{MIMO}{Multiple-Input Multiple-Output}

\begin{document}

\title{New LISA dynamics feedback control scheme: Common-mode isolation of test mass control and probes of test-mass acceleration}

\newcommand{\indice}[1]{{\scriptscriptstyle #1}}
\newcommand{\exposant}[1]{{\scriptscriptstyle #1}}
\newcommand{\myvec}[2]{\vec{#1}_\indice{#2}}
\newcommand{\myexpr}[3]{#1_\indice{#2}^\exposant{#3}}
% \DeclarePairedDelimiterX{\norm}[1]{\lVert}{\rVert}{#1}
\newcommand{\ddt}[2]{ \frac{d}{dt}\Bigg|_{\mathcal{#2}} \left[ #1 \right]}
\newcommand{\ddtddt}[2]{ \frac{d^2}{dt^2}\Bigg|_{\mathcal{#2}} \left[ #1 \right]}
\newcommand{\myhyperref}[1]{\hyperref[#1]{\ref{#1}}}
\newcommand{\identite}[1]{{\displaystyle \mathbb{1}_{\indice{#1}}}}
\newcommand{\myskew}[2][]{\@ifmtarg{#1}{\left[ #2 \right]^{\times}}{\left[ #2 \right]^{\times, #1}}}
\newcommand{\myat}[2][]{#1|_{#2}}
\newcommand{\timestentothe}[1]{\times 10^{#1}}
\newcommand{\msout}[1]{\text{\sout{\ensuremath{#1}}}}
\newcommand{\sensor}[1]{\exposant{\text{{#1}}}}
\renewcommand{\arraystretch}{1.5}

\author{Henri Inchausp\'e}
\email[Corresponding author: ]{inchauspe@tphys.uni-heidelberg.de}
\affiliation{Universit\'e Paris Cit\'e, CNRS, CNES, Astroparticule et Cosmologie, F-75013 Paris, France}
\affiliation{Institut f\"ur Theoretische Physik, Universit\"at Heidelberg, Philosophenweg 16, 69120 Heidelberg, Germany}
% \author{Henri Inchausp\'e}\altaffiliation[Current Address: ]{\addressitp}\affiliation{\addressapc}

\author{Martin Hewitson}
\affiliation{Albert-Einstein-Institut, Max-Planck-Institut f\"ur Gravitationsphysik und Leibniz Universit\"at Hannover,
Callinstra{\ss}e 38, 30167 Hannover, Germany}

\author{Orion Sauter}
\affiliation{Department of Mechanical and Aerospace Engineering, MAE-A, P.O. Box 116250, University of Florida, Gainesville, Florida 32611, USA}

\author{Peter Wass}
\affiliation{Department of Mechanical and Aerospace Engineering, MAE-A, P.O. Box 116250, University of Florida, Gainesville, Florida 32611, USA}

%% Abstract

\begin{abstract}

    The Drag-Free and Attitude Control System is a central element of LISA technology, ensuring the very high dynamic stability of spacecraft and test masses required in order to reach the  sensitivity that gravitational wave astronomy in space requires. Applying electrostatic forces on test-masses is unavoidable but should be restricted to the minimum necessary to keep the spacecraft-test masses system in place, while granting the optimal quality of test-mass free-fall. To realise this, we propose a new test-mass suspension scheme that applies forces and torques only in proportion to any differential test mass motion observed, and we demonstrate that the new scheme significantly mitigates the amount of suspension forces and torques needed to control the whole system. The mathematical method involved allows us to derive a new observable measuring the differential acceleration of test masses projected on the relevant sensitive axes, which will have important consequences for LISA data calibration, processing and analysis.

\end{abstract}

\maketitle

\section{Introduction}

The \gls{lisa} \cite{amaro-seoane_laser_2017-1} will detect gravitational waves from space in the $[10^{-4}\,\si{\hertz}$-$1\,\si{\hertz}]$ frequency band, opening a new window on the Universe and providing access to diverse astrophysical sources, including mergers of super-massive black hole binary systems, \glspl{emri}, galactic ultra-compact binaries \cite{klein_science_2016, marsat_exploring_2020} and stellar-mass black hole binaries (LIGO-Virgo-like sources) during their inspiral phase. LISA may also detect a stochastic cosmological gravitational wave signal\cite{caprini_cosmological_2018, caprini_science_2016, bartolo_science_2016}, which would have a significant impact on our understanding of the dynamics of the early Universe and fundamental physics. Such a detection will rely strongly on a deep understanding and knowledge of the instrumental noise in LISA.

The \gls{lisa} instrument is formed by a constellation of three spacecraft (S/C) placed at the vertices of a quasi-equilateral triangle orbiting the Sun. The gravitational wave detection principle reflects the usual picture of their effect on matter: a network of free particles is deformed by gravitational radiation passing through. Interferometry between free-falling spacecraft---representing free particles---is realized to measure the deformation. It follows that there are two immediate, essential challenges the LISA technology needs to address:

\begin{itemize}
    \item Direct Michelson-like interferometry is not possible for a $2.5$ million $\si{\kilo\metre}$ scale constellation in space, since unequal arm-lengths are imposed by orbital dynamics and laser power on-board is not sufficient for a light round-trip between spacecraft. Therefore a transponder-like scheme combined with post-processing synthesis of the measurements using time-delay interferometry is required \cite{tinto_time-delay_2020, vallisneri_geometric_2005}.
    \item Spacecraft are poor references of inertia due to external disturbances such as solar radiation pressure. Instead the spacecraft carry cubic Pt-Au alloy 1.92\,\si{\kilo\gram} test masses, shielded from the space environment and free-falling at the $\si{\femto\metre\per\second\squared}$ level \cite{armano_sub-femto-g_2016, armano_beyond_2018}. The spacecraft dynamics will be locked onto the test mass motion (with the help of the so-called \gls{dfacs} \cite{lisa_pathfinder_collaboration_lisa_2019}), or monitored and accounted for during the post-processing formation of synthetic interferometers.
\end{itemize}

\begin{figure}[ht]
\centering
\frame{\centerline{\includegraphics[scale=0.304, trim={0.1cm 0.0cm 0.2cm 0cm}, clip]{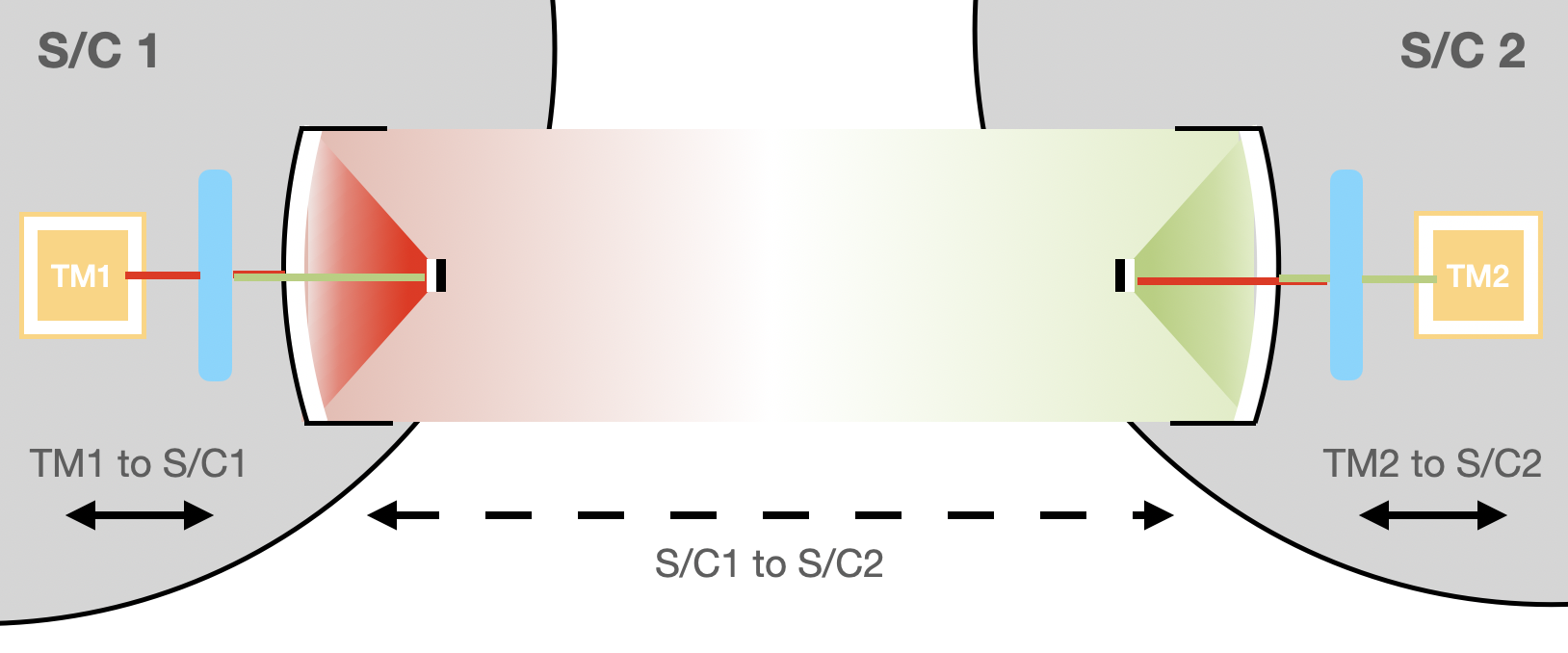}}}
\caption{Diagram of the long-range measurement breakdown. The test mass-to-test mass optical measurement is decomposed into three pieces, two local and one long-range (2.5 million of kilometres) in a transponder-like measurement scheme.}
\label{figure: Images/longrange}	
\end{figure}

This second challenge has been the object of a dedicated space mission, LISA Pathfinder \cite{armano_sub-femto-g_2016} \cite{armano_beyond_2018}, as a technological demonstrator of most of the space metrology subsystems \cite{armano_sensor_2021, lisa_pathfinder_collaboration_capacitive_2017} and critical technologies \cite{lisa_pathfinder_collaboration_precision_2018} on-board the LISA satellites, except for the long-range interferometry which could not be tested on a single spacecraft. In the LISA detection principle, no direct, long-range test mass-to-test mass optical measurement is possible, as already mentioned above. Instead, a transponder-like scheme is used (cf. Fig. \myhyperref{figure: Images/longrange}) where the test mass-to-test mass measurement is broken down into three consecutive measurements, that is, test mass-to-spacecraft (local), spacecraft-to-spaceraft (inter-spacecraft or long-range) and spacecraft-to-test mass (local). This necessity of decomposing the measurements introduces imperfections, most important of which, are optical misalignments which can lead to important cross-couplings, such as optical tilt-to-length couplings \cite{chwalla_optical_2020}, limiting the sensitivity of the instrument between $10$ and $100$ \si{\milli\hertz} if no hardware or post-processing corrections are undertaken. The noise introduced by such cross-couplings are driven by the spacecraft and telescope jitters and therefore the minimization of the latter is of critical importance.

The \gls{dfacs} is therefore a core subsystem of the LISA technology, which ensures a very high level of stability of the test masses inside the spacecraft in order to mitigate force gradients and, in general, measure couplings between the payload platform motion---the satellite jitter---and local and long-range optical path-lengths. A critical aspect of the problem is to realize an optimal decoupling between the so-called differential mode motion, driven by the relative motions and sensing of the test-masses, and the common-mode motion driven by the spacecraft, a common supporting platform to both test masses. Such isolation between common and differential modes will ensure minimal couplings between the noisy platform and the sub-picometer interferometry at aim.

Design of the actual LISA \gls{dfacs} is an on-going activity, with preliminary studies on a comprehensive, non-linear modeling of the system \cite{vidano_lisa_2020} recently published. A thorough analysis of LISA Pathfinder DFACS performance \cite{lisa_pathfinder_collaboration_lisa_2019, schleicher_-orbit_2018} has paved the way for a better understanding and optimization of LISA control, as well as demonstrating the reliability of closed-loop dynamics simulations for explaining stability and performance data. In this paper, we propose a novel \gls{dfacs} scheme which mitigates significantly necessary commanded forces on test masses---the so-called suspension forces and torques---in decoupling them from the spacecraft jitter. After introducing the reader to the \gls{dfacs} control principle and strategy, the article derives in Section \myhyperref{section: DFACS scheme} the new control coordinates which ensures decoupling of common and differential mode motions. In Section \myhyperref{section: simulation}, simulation experiments demonstrate the efficiency of this isolation scheme in ensuring that stability performance is left intact. Section \myhyperref{section: budget} presents the impact of this new scheme on LISA noise budget, with a focus on the mitigation of actuation crosstalk triggered by platform jitter. Finally Section \myhyperref{section: acceleration} presents a novel method of measuring test mass (differential) acceleration noise in-orbit making use of the new suspension scheme algebra.

\section{LISA Drag-Free and Attitude Control System}

\begin{figure}[h!]

\frame{\centerline{\includegraphics[scale=0.26, trim={0.0cm 1.0cm 0.0cm 0cm}, clip]{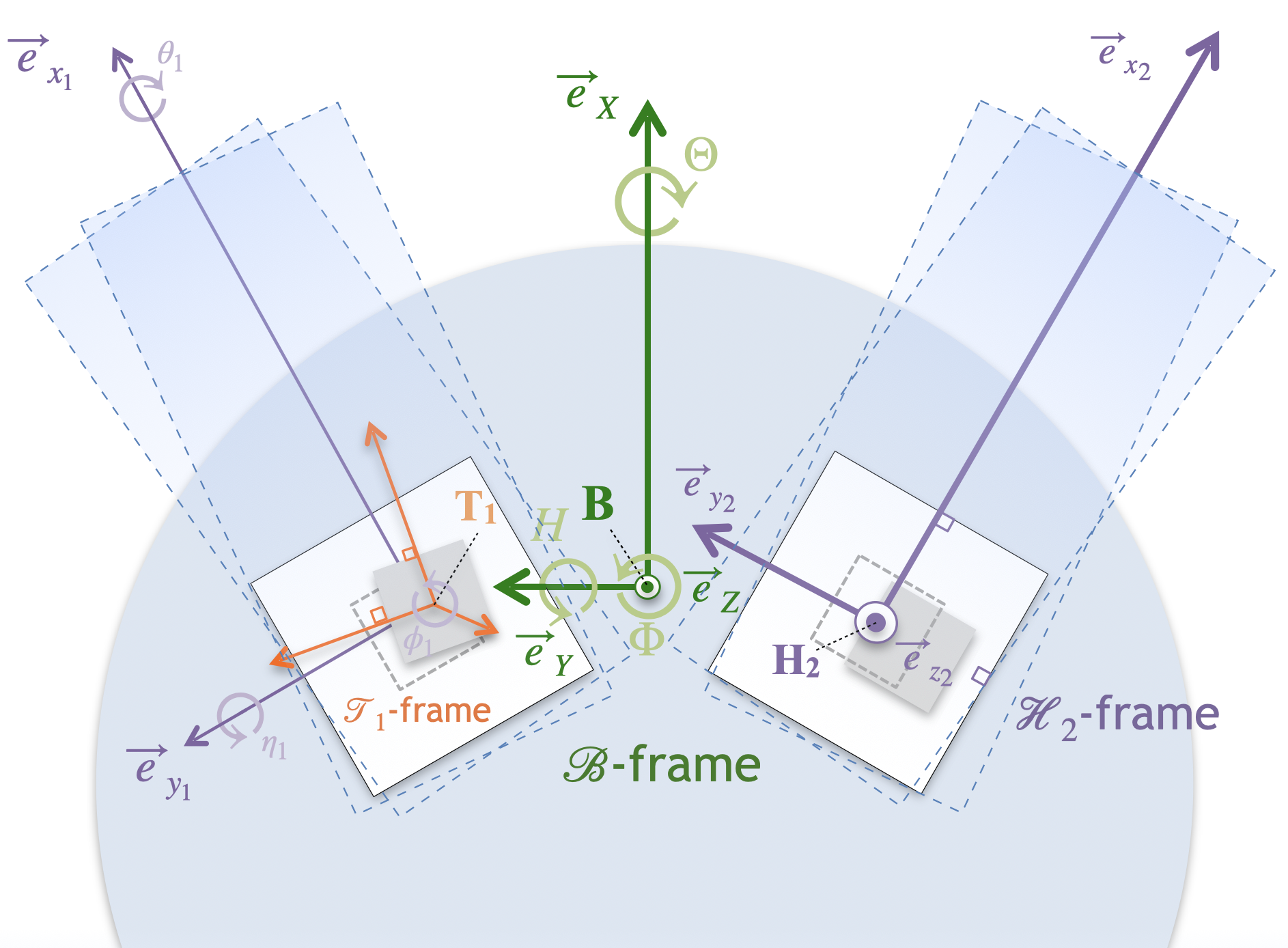}}}
\caption{Cross-section of the spacecraft geometry on the $XOY$ plane of the spacecraft body frame $\mathcal{B} = \{\myvec{e}{X}, \myvec{e}{Y}, \myvec{e}{Z}\}$. The $\mathcal{B}$-frame is located at the spacecraft centre-of-mass $B$, and its axes are defined so that $\myvec{e}{X}$ is aligned with the bisector between the two telescope axes in the standard, symmetric configuration (opening angle $\phi_{\text{m}} = 60 \si{\degree}$), $\myvec{e}{Z}$ is normal to the spacecraft solar panels and $\myvec{e}{Y}$ then completes the triad. Spacecraft attitude is encoded with cardan angles $[\Theta, H, \Phi]$ around $\mathcal{B}$-frame axes. Longitudinal motion $[x_{1/2}, y_{1/2}, z_{1/2}]$ of the test masses are tracked w.r.t. their respective housing frame $\mathcal{H}_\indice{1/2} = \{\myvec{e}{x_{1/2}}, \myvec{e}{y_{1/2}}, \myvec{e}{z_{1/2}}\}$, whose axes are set normal to housing inner walls and origins at housing geometrical centres. Test mass frames $\mathcal{T}_\indice{1/2}$ describe test mass orientation deviation $\vec{\alpha}_{\mathcal{T}_\indice{1/2}/\mathcal{H}_\indice{1/2}} = [\theta_\indice{1/2}, \eta_\indice{1/2}, \phi_\indice{1/2}]$ from their nominal orientation represented by the $\mathcal{H}_\indice{1/2}$ frames.}
\label{figure: Images/spacecraft}

\end{figure}

The DFACS feedback control strategy, ensuring longitudinal and angular stability of spacecraft and test masses, can be split into three sub-components:

\begin{itemize}
    \item {\it Drag-Free} control, which is used to lock the spacecraft longitudinal motion onto the much quieter test masses. Actuation from the micro-thruster system is used in order to compensate for stray forces and actuation applied to the spacecraft, ensuring that any stray forces must be corrected by forces applied on the spacecraft only. On the $XOY$ plane of the spacecraft (cf. Fig. \myhyperref{figure: Images/spacecraft}), which includes the two sensitive axes, {\it Drag-Free} control acts upon information from the local test-mass \gls{ifo}.
    \item {\it Attitude} control, which constrains the orientation of the spacecraft relative to the incoming laser wavefronts emitted from the distant spacecraft---hence locking the triangular constellation. It utilizes \gls{ldws} to orient the spacecraft w.r.t. the constellation and compensates stray, external torques with the micro-propulsion system. In addition, orbital constellation breathing, in which the opening angle between \glspl{mosa} (MOSA) can change by up to $\pm1^{\circ}$, is accounted for via a mechanism acting which introduces a fourth control degree of freedom (d.o.f.) (in a symmetric actuation configuration).
    \item {\it Suspension} control, which consists of the application of electrostatic forces on the test masses by applying voltages to the surrounding set of electrodes distributed over the inner surface of the \gls{grs} housings \cite{dolesi_gravitational_2003} \cite{lisa_pathfinder_collaboration_capacitive_2017}. Such forces are required to compensate any differential acceleration between the test masses that {\it Drag-Free} will not be able to correct by construction. The suspension force authority must be very limited in order to mitigate actuation noise and stray force gradients in the housing. Stray forces that accelerate the test masses (with the DC component mainly being driven by spacecraft and self-gravity) are expected to be low by design ($< 0.3\,\si{\nano\metre\per\second\squared}$ at DC) \cite{armano_constraints_2016}.
\end{itemize}

The overall 18 d.o.f. control scheme relies on three sensor sub-systems: the test-mass \gls{ifo} providing around 5\,\si{\pico\metre\persqrthz} measurement precision of longitudinal displacement along each of their $x$-axes, and 5\,\si{\nano\radian\persqrthz} angular displacement precision around axes orthogonal to the $x$-axis ($\eta$ and $\phi$); a capacitive-sensing system---\gls{grs} sensing---significantly less precise than the optical system, but available for all $6$ d.o.f. of each test mass, and providing roughly 1\,\si{\nano\metre\persqrthz} and 0.1\,\si{\micro\radian\persqrthz} test mass-to-housing displacement measurement precision \cite{lisa_pathfinder_collaboration_capacitive_2017}; and finally, the long-range \gls{ldws} measuring spacecraft attitude w.r.t. the received beams from the far spacecraft at the 0.2\,\si{\nano\radian\persqrthz} level (when accounting for telescope and optical bench imaging magnification factors). The reader will find a more accurate and quantitative listing of the sensing performance---the actual settings used in the simulations discussed---in Table \myhyperref{table: NoiseLevels} in Section \myhyperref{subsection: Simulation software and control laws}. Table \myhyperref{table: DFACS} summarizes the \gls{dfacs} configuration and the typical sensing-to-actuation mapping.

\begin{table}%[H] add [H] placement to break table across pages
\scriptsize
\caption{Example of a simple control scheme of LISA in a proposed, {\it nominal science mode}. For each control coordinate, the table lists the respective control type and actuator used, as well as the subsystem they are sensed with. Only the simple control scheme case (see Appendix \myhyperref{appendix: simple scheme}) is shown for readability. Capital letters are used for spacecraft coordinates, while lower case and indices are used for test mass coordinates. In the text, spacecraft and test mass coordinates may be labelled to specify either the sensor system which provides its measurement or the control coordinates scheme used.}
\label{table: DFACS}
\begin{ruledtabular}
\begin{tabular}{|c|c|c|c|c|c|}
\# & Coordinate & Sensor & Control Mode & Actuator & Command \\ \hline\hline
1 & $\hat{x}_\indice{1}$ & \gls{ifo} & Drag-Free & $\mu$-thrust & $F_{X}^{\text{drag-free}}$ \\
2 & $\hat{x}_\indice{2}$ & \gls{ifo} & Drag-Free & $\mu$-thrust & $F_{Y}^{\text{drag-free}}$ \\
3 & $\hat{z}_\indice{1}$ & \gls{grs} & Drag-Free & $\mu$-thrust & $F_{Z}^{\text{drag-free}}$ \\
4 & $\hat{\Theta}$ & \gls{ldws} & Attitude & $\mu$-thrust & $N_{X}^{\text{att}}$ \\
5 & $\hat{H}$ & \gls{ldws} & Attitude & $\mu$-thrust & $N_{Y}^{\text{att}}$ \\
6 & $\hat{\Phi}$ & \gls{ldws} & Attitude & $\mu$-thrust & $N_{Z}^{\text{att}}$ \\
7 & $\hat{y}_\indice{1}$ & \gls{grs} & Suspension & \gls{grs} & $F_{y_1}^{\text{sus}}$ \\
8 & $\hat{y}_\indice{2}$ & \gls{grs} & Suspension & \gls{grs} & $F_{y_2}^{\text{sus}}$ \\
9 & $\hat{z}_\indice{2}$ & \gls{grs} & Suspension & \gls{grs} & $F_{z_1}^{\text{sus}}$ / $F_{z_1}^{\text{sus}}$ \\
10 & $\hat{\theta}_\indice{1}$ & \gls{grs} & Suspension & \gls{grs} & $N_{x_1}^{\text{sus}}$ \\
11 & $\hat{\eta}_\indice{1}$ & \gls{ifo} & Suspension & \gls{grs} & $N_{y_1}^{\text{sus}}$ \\
12 & $\hat{\phi}_\indice{1}$ & \gls{ifo} & Suspension & \gls{grs} & $N_{z_1}^{\text{sus}}$ \\
13 & $\hat{\theta}_\indice{2}$ & \gls{grs} & Suspension & \gls{grs} & $N_{x_2}^{\text{sus}}$ \\
14 & $\hat{\eta}_\indice{2}$ & \gls{ifo} & Suspension & \gls{grs} & $N_{y_2}^{\text{sus}}$ \\
15 & $\hat{\phi}_\indice{2}$ & \gls{ifo} & Suspension & \gls{grs} & $N_{z_2}^{\text{sus}}$ \\
\end{tabular}
\end{ruledtabular}
\end{table}

\section{DFACS scheme optimization \& isolation of suspension control}
\label{section: DFACS scheme}

In this section the proposed, optimized control coordinates allowing for an isolation of suspension control from spacecraft jitter are derived. In line with the notation of table \myhyperref{table: DFACS}, we write such optimal coordinates $[ \hat{x}_\indice{1}^\exposant{\text{opt}}, \allowbreak  \hat{x}_\indice{2}^\exposant{\text{opt}}, \allowbreak  \hat{z}_\indice{1}^\exposant{\text{opt}}, \allowbreak  \hat{\Theta}^\exposant{\text{opt}}, \allowbreak  \hat{H}^\exposant{\text{opt}}, \allowbreak  \hat{\Phi}^\exposant{\text{opt}}, \allowbreak  \hat{y}_\indice{1}^\exposant{\text{opt}}, \allowbreak  \hat{y}_\indice{2}^\exposant{\text{opt}}, \allowbreak  \hat{z}_\indice{2}^\exposant{\text{opt}}, \allowbreak  \hat{\theta}_\indice{1}^\exposant{\text{opt}}, \allowbreak  \hat{\eta}_\indice{1}^\exposant{\text{opt}}, \allowbreak  \hat{\phi}_\indice{1}^\exposant{\text{opt}}, \allowbreak  \hat{\theta}_\indice{2}^\exposant{\text{opt}}, \allowbreak  \hat{\eta}_\indice{2}^\exposant{\text{opt}}, \allowbreak  \hat{\phi}_\indice{2}^\exposant{\text{opt}}]$, as opposed to the simpler scheme coordinates $[ \hat{x}_\indice{1}^\exposant{\text{sim}}, \allowbreak  \hat{x}_\indice{2}^\exposant{\text{sim}}, \allowbreak  \hat{z}_\indice{1}^\exposant{\text{sim}}, \allowbreak  \hat{\Theta}^\exposant{\text{sim}}, \allowbreak  \hat{H}^\exposant{\text{sim}}, \allowbreak  \hat{\Phi}^\exposant{\text{sim}}, \allowbreak  \hat{y}_\indice{1}^\exposant{\text{sim}}, \allowbreak  \hat{y}_\indice{2}^\exposant{\text{sim}}, \allowbreak  \hat{z}_\indice{2}^\exposant{\text{sim}}, \allowbreak  \hat{\theta}_\indice{1}^\exposant{\text{sim}}, \allowbreak  \hat{\eta}_\indice{1}^\exposant{\text{sim}}, \allowbreak  \hat{\phi}_\indice{1}^\exposant{\text{sim}}, \allowbreak  \hat{\theta}_\indice{2}^\exposant{\text{sim}}, \allowbreak  \hat{\eta}_\indice{2}^\exposant{\text{sim}}, \allowbreak  \hat{\phi}_\indice{2}^\exposant{\text{sim}}]$ (see appendix \myhyperref{appendix: simple scheme} for details) mentioned throughout and studied for comparison.

\subsection{Drag-Free and common-mode correction}

An optimal drag-free control is designed so that any observed test-mass displacement induced by an acceleration of the spacecraft w.r.t. its local inertial frame is only corrected through actuation thrust on the spacecraft itself: forces and torques on test masses will only arise in proportion to any sensed differential motion of the two test masses. Complying with such a philosophy, one can build from the observed test mass displacements---as observed by the local interferometers---coordinates that we call {\it common-mode} coordinates and which provide the best measurement of the spacecraft acceleration w.r.t. inertial space.

Based on the spacecraft geometry and the opening angle between the two \gls{mosa}'s $\phi_{\text{m}}$, we define the common-mode coordinates, $\{\myvec{e}{X}, \myvec{e}{Y}, \myvec{e}{Z}\}$ in terms of the test mass coordinates $\{\myvec{e}{x_{1/2}}, \myvec{e}{y_{1/2}}, \myvec{e}{z_{1/2}}\}$, both reference frames shown and detailed in Figure \myhyperref{figure: Images/spacecraft}:
\begin{align}
\label{eq: H_2_B_vectors}
    & \myvec{e}{X} = \frac{\myvec{e}{x_1} + \myvec{e}{x_2}}{\sqrt{2 + 2 \cos{\phi_{\text{m}}}}} = \frac{\myvec{e}{x_1} + \myvec{e}{x_2}}{\sqrt{3}} \nonumber \\
    & \myvec{e}{Y} = \frac{\myvec{e}{x_1} - \myvec{e}{x_2}}{\sqrt{2 \cos{\phi_{\text{m}}}}} = \myvec{e}{x_1} - \myvec{e}{x_2} \\
    & \myvec{e}{Z} = \myvec{e}{z_1} = \myvec{e}{z_2} \nonumber.
\end{align}
where setting the opening angle between telescopes at $\phi_{\text{m}} = 60 \si{\degree}$ introduces the factors $\tfrac{1}{\sqrt{3}}$ and $1.0$ for the $\myvec{e}{X}$ and $\myvec{e}{Y}$ drag-free directions respectively.
These common-mode coordinates are tracked by drag-free control using the sensing channels listed in Table \ref{table: NoiseLevels}. Consequently {\it drag-Free} control requested thrusts on the spacecraft will be proportional to these common-mode combinations, and are defined as:

\begin{empheq}[box=\mymath]{align}
\label{eq: DragFreeScheme}
    & F_{X}^{\text{drag-free}} \propto \hat{x}_\indice{1}^\exposant{\text{opt}} \equiv \frac{x_1^\sensor{ifo} + x_2^\sensor{ifo}}{\sqrt{2 + 2 \cos{\phi_{\text{m}}}}} = \frac{x_1^\sensor{ifo} + x_2^\sensor{ifo}}{\sqrt{3}} \nonumber \\
    & F_{Y}^{\text{drag-free}} \propto \hat{x}_\indice{2}^\exposant{\text{opt}} \equiv \frac{x_1^\sensor{ifo} - x_2^\sensor{ifo}}{\sqrt{2 \cos{\phi_{\text{m}}}}} = x_1^\sensor{ifo} - x_2^\sensor{ifo} \\
    & F_{Z}^{\text{drag-free}} \propto \hat{z}_\indice{1}^\exposant{\text{opt}} \equiv \frac{z_1^\sensor{grs} + z_2^\sensor{grs}}{2}. \nonumber
\end{empheq}
where one uses the redundancy of the $z_\indice{1-2}^\sensor{grs}$ measurements for averaging and picking the common-mode motion along $\myvec{e}{Z}$. In this new scheme, the $x_1^\sensor{ifo}$ and $x_2^\sensor{ifo}$ remain drag-free controlled effectively, and the test masses apparent motion in their housings are nulled by commanded thrust, essentially compensating for the spacecraft motion jitter.

In equation \myhyperref{eq: DragFreeScheme} we introduce test-masses coordinates labelled with the sensor system used to measure them in flight. We refer the reader to table \myhyperref{table: DFACS} which lists the dynamical control coordinates and the associated sensors and actuators used for control. Throughout the paper, for notation simplicity, labels will be dropped during mathematical demonstrations, and will be reintroduced at the final stage of the derivation only.

\subsection{Attitude control}

The angular dynamics of the spacecraft are locked on the \gls{ldws} sensors, which set the constellation reference frame for the spacecraft to rotate with in order to ensure that the telescopes are pointing towards the distant spacecraft. The incident angles $[\eta_1^\sensor{ldws}, \phi_1^\sensor{ldws}]$ and $[\eta_2^\sensor{ldws}, \phi_2^\sensor{ldws}]$ of the two distant laser beams as received by the local spacecraft telescopes yields a measurement of its attitude relatively to the quasi-inertial (for timescale shorter than its annual rotation) constellation frame. The \gls{ldws} provides sub-nanoradian attitude sensing precision, sufficient for use in the spacecraft angular jitter compensation loop. Consequently, in this scheme, attitude control is both used for the spacecraft to track a reference orientation determined by the constellation orbit, as well as for angular jitter mitigation---or stated differently as angular drag-free control.

From the four \gls{ldws} angles $[\eta_1^\sensor{ldws}, \phi_1^\sensor{ldws}, \eta_2^\sensor{ldws}, \phi_2^\sensor{ldws}]$ and the geometry of the spacecraft, one can estimate the spacecraft attitude w.r.t. its target orientation frame. The Cardan angles $\Theta$, $H$ and $\Phi$ \cite{diebel_representing_2006}
are determined by Equation \ref{eq: AttDetMatrix}, derived from spacecraft and \gls{mosa} geometry:
\begin{align}
\label{eq: SusAxes_2_B}
    & \myvec{e}{X} = \frac{\myvec{e}{y_2} - \myvec{e}{y_1}}{\sqrt{2 \cos{\phi_{\text{m}}}}} = \myvec{e}{y_2} - \myvec{e}{y_1} \nonumber \\
    & \myvec{e}{Y} = \frac{\myvec{e}{y_1} + \myvec{e}{y_2}}{\sqrt{2 + 2 \cos{\phi_{\text{m}}}}} = \frac{\myvec{e}{y_1} + \myvec{e}{y_2}}{\sqrt{3}} \\
    & \myvec{e}{Z} = \myvec{e}{z_1} = \myvec{e}{z_2}. \nonumber
\end{align}
and assuming an opening angle of $60 \si{\degree}$.
\begin{align}
\label{eq: AttDetMatrix}
    \begin{bmatrix}
    & \Theta^\sensor{ldws} \\
    & H^\sensor{ldws} \\
    & \Phi^\sensor{ldws}
    \end{bmatrix}
    =
    \begin{pmatrix}
        & 0.0 & -1.0 & 0.0 & 1.0 \\
        & 0.0 & -\tfrac{1}{\sqrt{3}} & 0.0 & -\tfrac{1}{\sqrt{3}} \\
        & -0.5 & 0.0 & -0.5 & 0.0
    \end{pmatrix}
    \begin{bmatrix}
    & \phi_1^\sensor{ldws} \\
    & \eta_1^\sensor{ldws} \\
    & \phi_2^\sensor{ldws} \\
    & \eta_2^\sensor{ldws}
    \end{bmatrix}.
\end{align}
Hence the error signals for attitude control---those triggering spacecraft angular thrust commands $N^{\text{att}}$ ---are defined as

\begin{empheq}[box=\mymath]{align}
\label{eq: AttitudeScheme}
    & N_{X}^{\text{att}} \propto \hat{\Theta}^\exposant{\text{opt}} \equiv \Theta^\sensor{ldws} = \eta_2^\sensor{ldws} - \eta_1^\sensor{ldws} \nonumber \\
    & N_{Y}^{\text{att}} \propto \hat{H}^\exposant{\text{opt}} \equiv H^\sensor{ldws} = -\frac{1}{\sqrt{3}} \left( \eta_1^\sensor{ldws} + \eta_2^\sensor{ldws} \right) \\
    & N_{Z}^{\text{att}} \propto \hat{\Phi}^\exposant{\text{opt}} \equiv \Phi^\sensor{ldws} = -\frac{1}{2} \left( \phi_1^\sensor{ldws} + \phi_2^\sensor{ldws} \right) \nonumber
\end{empheq}

\subsection{Suspension and differential mode: Longitudinal isolation}
\label{subsection: Suspension and differential mode: Longitudinal isolation}

In order to optimize the decoupling between common-mode and differential-mode test mass dynamics, one has to ensure that the suspension control is locked on the differential displacement of the two test masses, $\overrightarrow{\text{T}_\indice{1} \text{T}_\indice{2}}$
\begin{align}\label{eq: tm_rel_pos}
    \overrightarrow{\text{T}_\indice{1} \text{T}_\indice{2}} &= \overrightarrow{\text{T}_\indice{1} \text{H}_\indice{1}} + \overrightarrow{\text{H}_\indice{1} \text{H}_\indice{2}} + \overrightarrow{\text{H}_\indice{2} \text{T}_\indice{2}} \\ \nonumber
    &= \myvec{r}{\text{T}_\indice{2}/\text{H}_\indice{2}} - \myvec{r}{\text{T}_\indice{1}/\text{H}_\indice{1}} + \overrightarrow{\text{H}_\indice{1} \text{H}_\indice{2}}.
\end{align}
where $\myvec{r}{T_\indice{k}/H_\indice{k}}$ is the test-mass $k$ displacement vector within its respective housing, observed by interferometers and capacitive sensors:
\begin{align}
     & \myvec{r}{\text{T}_\indice{1}/\text{H}_\indice{1}} = x_1^\sensor{ifo} \myvec{e}{x_1} + y_1^\sensor{grs} \myvec{e}{y_1} + z_1^\sensor{grs} \myvec{e}{z_1} \nonumber \\
     & \myvec{r}{\text{T}_\indice{2}/\text{H}_\indice{2}} = x_2^\sensor{ifo} \myvec{e}{x_2} + y_2^\sensor{grs} \myvec{e}{y_2} + z_2^\sensor{grs} \myvec{e}{z_2},
\label{eq: observed position vectors}
\end{align}
and $\overrightarrow{\text{H}_\indice{1} \text{H}_\indice{2}}$ is the nominal static offset between the test masses. Small changes in the attitude of the spacecraft $\myvec{\alpha}{\mathcal{B}/\mathcal{B^*}}$ relative to the target frame $\mathcal{B^*}$ introduce contributions from levers to the apparent differential displacement $\Delta \overrightarrow{\text{T}_\indice{1} \text{T}_\indice{2}}$, for which one needs to account.

\begin{align}
    \Delta \overrightarrow{\text{T}_\indice{1} \text{T}_\indice{2}} &= \Delta \vec{r}_{\text{diff}} + \Delta \vec{r}_{\text{lever}} \nonumber\\
    &=\left[ \myvec{r}{\text{T}_\indice{2}/\text{H}_\indice{2}} - \myvec{r}{\text{T}_\indice{1}/\text{H}_\indice{1}} \right] + \left[ \myvec{\alpha}{\mathcal{B}/\mathcal{B}^*} \times \overrightarrow{\text{H}_\indice{1} \text{H}_\indice{2}} \right]
\end{align}

Isolating the translational suspension control from spacecraft jitter will then consist in soliciting the electrostatic feedback along axes $\myvec{e}{y_1}$, $\myvec{e}{z_1}$, $\myvec{e}{y_2}$ and $\myvec{e}{z_1}$ to null the quantity $\Delta \overrightarrow{\text{T}_\indice{1} \text{T}_\indice{2}}$. Therefore, one needs to project $\Delta \overrightarrow{\text{T}_\indice{1} \text{T}_\indice{2}}$ along those axes, and in doing so, express all the vector quantities in a common coordinate system fixed in the S/C frame:
\begin{align}
\label{eq: B2Hframe}
     & \myvec{e}{x_1} = \cos{\tfrac{\phi_\indice{\text{m}}}{2}} \myvec{e}{X} + \sin{\tfrac{\phi_\indice{\text{m}}}{2}} \myvec{e}{Y} \nonumber
     \\
     & \myvec{e}{y_1} = -\sin{\tfrac{\phi_\indice{\text{m}}}{2}} \myvec{e}{X} + \cos{\tfrac{\phi_\indice{\text{m}}}{2}} \myvec{e}{Y} \nonumber
     \\
     \\
     & \myvec{e}{x_2} = \cos{\tfrac{\phi_\indice{\text{m}}}{2}} \myvec{e}{X} - \sin{\tfrac{\phi_\indice{\text{m}}}{2}} \myvec{e}{Y} \nonumber
     \\
     & \myvec{e}{y_2} = \sin{\tfrac{\phi_\indice{\text{m}}}{2}} \myvec{e}{X} + \cos{\tfrac{\phi_\indice{\text{m}}}{2}} \myvec{e}{Y} \nonumber
\end{align}

Using this basis and treating the differential test mass displacement first, we can write:
\begin{align}
\label{eq: diff_tm_pos}
     & \Delta \vec{r}_{\text{diff}} =
     \begin{matrix}
      & & \left[ (x_2 - x_1) \cos{\tfrac{\phi_\indice{\text{m}}}{2}} + (y_2 + y_1) \sin{\tfrac{\phi_\indice{\text{m}}}{2}} \right] & \myvec{e}{X} \\
      & + & \left[ (-x_1 - x_2) \sin{\tfrac{\phi_\indice{\text{m}}}{2}} + (y_2 - y_1) \cos{\tfrac{\phi_\indice{\text{m}}}{2}} \right] & \myvec{e}{Y} \\
      & + & \left[ z_2 - z_1 \right] & \myvec{e}{Z}.     
     \end{matrix}
\end{align}
The suspension forces are generated by electrostatic actuation applied by the same set of capacitors used for position sensing. Therefore, the suspension is performed in the housing reference frames, along the $y$ and $z$ axes since no force shall be applied on the $x$ directions, the dimension along which the test masses must be free-falling. Hence one needs to project Equation\,(\myhyperref{eq: diff_tm_pos}) along suspension axes $y_1$, $y_2$, $z_1$ and $z_2$. 

Using Equation\,(\myhyperref{eq: AttDetMatrix}) and considering an opening angle of $60 \si{\degree}$, one finds after further expansion the following projection:
\begin{align}
\label{eq: diff_tm_pos_proj}
     & \Delta \vec{r}_{\text{diff}} =
     \begin{matrix}
         & & \left[ -\frac{2}{\sqrt{3}} x_2 + \frac{1}{\sqrt{3}} x_1 - y_1 \right] & \myvec{e}{y_1} \\
         & + & \left[ \frac{1}{\sqrt{3}} x_2 - \frac{2}{\sqrt{3}} x_1 + y_2 \right] & \myvec{e}{y_2} \\
         & + & \left[ z_2 - z_1 \right] & \frac{\myvec{e}{z_1} + \myvec{e}{z_2}}{2}. 
     \end{matrix}
\end{align}

At this stage, that is, locking the suspension scheme on $\Delta \vec{r}_{\text{diff}}$ only as in Equation\,(\myhyperref{eq: diff_tm_pos_proj}) still lets angular jitter contribution through from various levers. While the common-mode projection of such levers in the two housings will be invisible to a suspension locked on $\Delta \vec{r}_{\text{diff}}$ as in Equation\,(\myhyperref{eq: diff_tm_pos_proj})---although seen and corrected by drag-free control---the differential component will be interpreted as an apparent translational drift between test masses. Then, it is required to subtract those terms from suspension control. Formally, accounting for those levers consists of considering the change of the relative position between test masses from the perspective of a rotating reference frame, hence forcing suspension to disregard apparent, differential motion arising from system of coordinate variations.

Turning to the lever arm effect, according to Equation\,(\myhyperref{eq: tm_rel_pos}), when the S/C rotates, the lever $\overrightarrow{\text{H}_\indice{1} \text{H}_\indice{2}}$ between the two test-mass nominal positions then generates an apparent, {\bfseries differential motion} between test masses:

\begin{equation}
    \Delta \vec{r}_{\text{lever}} = \myvec{\alpha}{\mathcal{B}/\mathcal{B}^*} \times \overrightarrow{\text{H}_\indice{1} \text{H}_\indice{2}} = \myvec{\alpha}{\mathcal{B}/\mathcal{B}^*} \times \Delta \myvec{r}{H2/H1},
\end{equation}
which expressed in the $\mathcal{B}$-frame gives:
\begin{align}
\label{eq: diff_lever}
     & \Delta \vec{r}_{\text{lever}} =
     \begin{matrix}
      & & \left( H\Delta z_\indice{\text{H}_\indice{2}/\text{H}_\indice{1}} - \Phi \Delta y_\indice{\text{H}_\indice{2}/\text{H}_\indice{1}} \right) & \myvec{e}{X} \\
      & + & \left( \Phi \Delta x_\indice{\text{H}_\indice{2}/\text{H}_\indice{1}} - \Theta \Delta z_\indice{\text{H}_\indice{2}/\text{H}_\indice{1}} \right) & \myvec{e}{Y} \\
      & + & \left( \Theta \Delta y_\indice{\text{H}_\indice{2}/\text{H}_\indice{1}} - H \Delta x_\indice{\text{H}_\indice{2}/\text{H}_\indice{1}} \right) & \myvec{e}{Z}.    
     \end{matrix}
\end{align}
This latter expression needs to be projected along $\myvec{e}{y_1}$, $\myvec{e}{y_2}$, $\myvec{e}{z_1}$ and $\myvec{e}{z_2}$. Using Equation\,(\myhyperref{eq: SusAxes_2_B}), substituting in Equation\,(\myhyperref{eq: diff_lever}), and simplifying hereafter the notation of the lever vector coordinates for clarity as $[\Delta x, \Delta y, \Delta z]$, one obtains:
\begin{align}
\label{eq: diff_lever_proj}
     & \Delta \vec{r}_{\text{lever}} =
     \begin{matrix}
      & & \left( H\Delta z - \Phi \Delta y \right) & \left( \myvec{e}{y_2} - \myvec{e}{y_1} \right) \\
      & + & \left( \Phi \Delta x - \Theta \Delta z \right) & \frac{\myvec{e}{y_1} + \myvec{e}{y_2}}{\sqrt{3}} \\
      & + & \left( \Theta \Delta y - H \Delta x \right) & \frac{\myvec{e}{z_1} + \myvec{e}{z_2}}{2}.
     \end{matrix}
\end{align}
The lever-arm component of the suspension force $F^{\text{lever}}$ along housing $y$ and $z$ axes must be triggered from observed test-mass differential motion as projected along the housing reference frame set of axes:
\begin{align}
    & F_{y_{1\text{-}2}}^{\text{lever}} \propto \Delta \vec{r}_{\text{lever}} \cdot \myvec{e}{y_{1\text{-}2}} \nonumber \\
    & F_{z_{1\text{-}2}}^{\text{lever}} \propto \Delta \vec{r}_{\text{lever}} \cdot \myvec{e}{z_{1\text{-}2}}, \nonumber
\end{align}
which from Equation\,(\myhyperref{eq: diff_lever_proj}) and after further expansion yields finally:
\begin{align}
    & F_{y_{1\text{-}2}}^{\text{lever}} \propto -\frac{\Delta z}{\sqrt{3}} \Theta \mp \Delta z H + \left( \frac{\Delta x}{\sqrt{3}} \pm \Delta y \right) \Phi \nonumber \\
    & F_{z_{1\text{-}2}}^{\text{lever}} \propto \mp \Delta y \Theta \pm \Delta x H. \nonumber \\
\end{align}
Finally, combining translational, differential motion and lever contributions, one arrives at the following suspension scheme:
\begin{empheq}[box=\mymath]{align}
\label{eq: scheme_sus_long}
     & F_{y_1}^{\text{sus}} \propto
     \hat{y}_\indice{1}^\exposant{\text{opt}} \equiv \begin{matrix}
         -\frac{1}{\sqrt{3}} x_1^\sensor{ifo} + \frac{2}{\sqrt{3}} x_2^\sensor{ifo} + y_1^\sensor{grs} \\
         + \frac{\Delta z}{\sqrt{3}} \Theta^\sensor{ldws} + \Delta z H^\sensor{ldws} \\
         - \left( \frac{\Delta x}{\sqrt{3}} + \Delta y \right) \Phi^\sensor{ldws}
     \end{matrix} \nonumber \\[0.2in]
     & F_{y_2}^{\text{sus}} \propto \hat{y}_\indice{2}^\exposant{\text{opt}} \equiv
     \begin{matrix}
         -\frac{2}{\sqrt{3}} x_1^\sensor{ifo} + \frac{1}{\sqrt{3}} x_2^\sensor{ifo} + y_2^\sensor{grs} \\
         -\frac{\Delta z}{\sqrt{3}} \Theta^\sensor{ldws} + \Delta z H^\sensor{ldws} \\
         + \left( \frac{\Delta x}{\sqrt{3}} - \Delta y \right) \Phi^\sensor{ldws}
     \end{matrix} \\[0.2in]
     & F_{z_1}^{\text{sus}} \propto \hat{z}_\indice{1}^\exposant{\text{opt}} \equiv
     \begin{matrix}
         z_1^\sensor{grs} - z_2^\sensor{grs} \\
         - \Delta y \Theta^\sensor{ldws} + \Delta x H^\sensor{ldws}
     \end{matrix} \nonumber \\[0.2in]
     & F_{z_2}^{\text{sus}} \propto \hat{z}_\indice{2}^\exposant{\text{opt}} \equiv
     \begin{matrix}
         z_2^\sensor{grs} - z_1^\sensor{grs} \\
         + \Delta y \Theta^\sensor{ldws} - \Delta x H^\sensor{ldws}
     \end{matrix} \nonumber
\end{empheq}

\subsection{Suspension and differential mode: Angular isolation}
An optimal suspension must also minimize coupling with angular jitter of the spacecraft. However, one must not entirely decouple angular suspension control from spacecraft attitude control, since the test mass rotation must follow the LISA constellation orbits around the solar system center. An optimal strategy consists of defining a common-mode angular control of the TMs based on the difference between the S/C attitude measurement with long-arm DWS and common-mode TM-DWS.

Making use of Equation\,(\myhyperref{eq: B2Hframe}) and the geometry of the spacecraft, one can express in a common coordinate system the local, test-mass DWS outputs and the long-range DWS outputs. Expressed in the S/C body frame $\mathcal{B}$, the common-mode suspension d.o.f written in terms of the sensing channels in Table\,\ref{table: NoiseLevels} would then take the form:
\begin{align}
    \theta_{\text{sus, c}} = & \frac{1}{2} \left[ \left( \theta_1 + \theta_2 \right) \cos{\sfrac{\phi_{\text{m}}}{2}} + \left( \eta_2 - \eta_1 \right) \sin{\sfrac{\phi_{\text{m}}}{2}} \right] \nonumber \\ & - \Theta \nonumber \\
    \eta_{\text{sus, c}} = & \frac{1}{2} \left[ \left( \theta_1 - \theta_2 \right) \sin{\sfrac{\phi_{\text{m}}}{2}} + \left( \eta_1 + \eta_2 \right) \cos{\sfrac{\phi_{\text{m}}}{2}} \right] \nonumber \\ & - H \\ 
    \phi_{\text{sus, c}} = & \frac{\phi_1 + \phi_2}{2} - \Phi, \nonumber   
\end{align}
which with a $60 \si{\degree}$ telescope angle reduces to:
\begin{align}
    & \theta_{\text{sus, c}} = \frac{\sqrt{3}}{4} \left( \theta_1 + \theta_2 \right) + \frac{1}{4} \left( \eta_2 - \eta_1 \right) - \Theta \nonumber \\
    & \eta_{\text{sus, c}} = \frac{1}{4} \left( \theta_1 - \theta_2 \right) + \frac{\sqrt{3}}{4} \left( \eta_1 + \eta_2 \right) - H \\ 
    & \phi_{\text{sus, c}} = \frac{\phi_1 + \phi_2}{2} - \Phi. \nonumber
\end{align}

While it is necessary to combine measured angles in a common coordinate system, the suspension torques $N^{\text{cmd}}$ are applied along coordinates fixed in the housing frames. Hence, it is required to project the common-mode coordinates onto the housing axes $\myvec{e}{x_1}$, $\myvec{e}{y_1}$, $\myvec{e}{z_1}$ and $\myvec{e}{x_2}$, $\myvec{e}{y_2}$, $\myvec{e}{z_2}$, which is performed invoking the appropriate rotation matrices:
\begin{align}
    \begin{bmatrix}
    & N_{x_{1\text{-}2}}^{\text{cmd}} \\
    & N_{y_{1\text{-}2}}^{\text{cmd}} \\
    & N_{z_{1\text{-}2}}^{\text{cmd}}
    \end{bmatrix}
    \propto
    \begin{pmatrix}
        & \sfrac{\sqrt{3}}{2} & \pm \sfrac{1}{2} & 0.0 \\
        & \mp \sfrac{1}{2} & \sfrac{\sqrt{3}}{2} & 0.0 \\
        & 0.0 & 0.0 & 1.0
    \end{pmatrix}
    \begin{bmatrix}
    & \theta_{\text{sus, c}}  \\
    & \eta_{\text{sus, c}}  \\
    & \phi_{\text{sus, c}} 
    \end{bmatrix},
\end{align}
again setting the opening angle to $60 \si{\degree}$. After further expansion of the common mode, one obtains the common-mode torque components $N^{\text{cmd,c}}$:
\begin{align}
    & N_{x_1}^{\text{cmd,c}} = \tfrac{1}{2} \theta_1 + \tfrac{1}{4} \theta_2 + \tfrac{\sqrt{3}}{4} \eta_2 - \tfrac{\sqrt{3}}{2} \Theta - \tfrac{1}{2} H
    \nonumber \\
    & N_{y_1}^{\text{cmd,c}} = \tfrac{1}{2} \eta_1 + \tfrac{1}{4} \eta_2  - \tfrac{\sqrt{3}}{4} \theta_2 + \tfrac{1}{2} \Theta - \tfrac{\sqrt{3}}{2} H
    \\[0.2in]
    & N_{x_2}^{\text{cmd,c}} = \tfrac{1}{2} \theta_2 + \tfrac{1}{4} \theta_1 - \tfrac{\sqrt{3}}{4} \eta_1 - \tfrac{\sqrt{3}}{2} \Theta + \tfrac{1}{2} H
    \nonumber \\
    & N_{y_2}^{\text{cmd,c}} = \tfrac{1}{2} \eta_2 + \tfrac{1}{4} \eta_1 + \tfrac{\sqrt{3}}{4} \theta_1 - \tfrac{1}{2} \Theta - \tfrac{\sqrt{3}}{2} H
\end{align}
with $\Theta$, $H$ and $\Phi$ taken out of attitude determination block in Equation\, (\myhyperref{eq: AttDetMatrix}).

For the differential mode, we proceed in a similar manner. The test mass DWS channels are combined after being expressed in coordinates fixed in the $\mathcal{B}$ frame:
\begin{align}
    & \theta_{\text{sus, d}} = \left( \theta_2 - \theta_1 \right) \cos{\tfrac{\phi}{2}} + \left( \theta_2 + \theta_1 \right) \sin{\tfrac{\phi}{2}}
    \nonumber \\
    & \eta_{\text{sus, d}} = - \left( \theta_2 + \theta_1 \right) \sin{\tfrac{\phi}{2}} + \left( \eta_2 - \eta_1 \right) \cos{\tfrac{\phi}{2}}
    \\
    & \phi_{\text{sus, d}} = \phi_2 - \phi_1. \nonumber
\end{align}
Applying the rotation matrices necessary to find the commanded torques applied in their respective housing frames and setting $\phi = 60 \si{\degree}$, this simplifies to give the differential mode torques $N^{\text{cmd,d}}$:

\begin{equation}
\left\{
\begin{aligned}
    & N_{x_1}^{\text{cmd,d}} = -\tfrac{1}{2} \theta_1 + \tfrac{1}{4} \theta_2 + \tfrac{\sqrt{3}}{4} \eta_2
    \nonumber \\
    & N_{y_1}^{\text{cmd,d}} = -\tfrac{1}{2}\eta_1 + \tfrac{1}{4} \eta_2 - \tfrac{\sqrt{3}}{4} \theta_2
    \\
    & N_{z_1}^{\text{cmd,d}} = \tfrac{1}{2} \phi_1 - \tfrac{1}{2} \phi_2
    \nonumber
\end{aligned}
\right.
\end{equation}

\begin{equation}
\left\{
\begin{aligned}
    & N_{x_2}^{\text{cmd,d}} = \tfrac{1}{2} \theta_2 - \tfrac{1}{4} \theta_1 + \tfrac{\sqrt{3}}{4} \eta_1
    \nonumber \\
    & N_{y_2}^{\text{cmd,d}} = \tfrac{1}{2} \eta_2 - \tfrac{1}{4} \eta_1 - \tfrac{\sqrt{3}}{4} \theta_1
    \\
    & N_{z_2}^{\text{cmd,d}} = \tfrac{1}{2} \phi_2 - \tfrac{1}{2} \phi_1,
    \nonumber
\end{aligned}
\right.
\end{equation}

which we combine with common mode control as follows, to get the total, net suspension torque:

\begin{align}
    & N_\indice{x_1, y_1, z_1}^{\text{cmd}} = N_\indice{x_1, y_1, z_1}^{\text{cmd,c}} - N_\indice{x_1, y_1, z_1}^{\text{cmd,d}} \\
    & N_\indice{x_2, y_2, z_2}^{\text{cmd}} = N_\indice{x_2, y_2, z_2}^{\text{cmd,c}} + N_\indice{x_2, y_2, z_2}^{\text{cmd,d}}.
\end{align}
Note the $\pm$ signs before contributions from the differential mode which differ between test mass 1 and 2 for geometrical reasons: correcting the differential ($2$ - $1$) channel implies applying forces on test masses along opposite directions. Simplifying further yields the final result:

\begin{empheq}[box=\mymath]{align}
\label{eq: scheme_sus_angle_tm}
    & N_{x_1}^{\text{sus}} \propto \hat{\theta}_\indice{1}^\exposant{\text{opt}} \equiv \theta_1^\sensor{grs} + \tfrac{\sqrt{3}}{2} \Theta^\sensor{ldws} + \tfrac{1}{2} H^\sensor{ldws}
    \nonumber \\
    & N_{y_1}^{\text{sus}} \propto \hat{\eta}_\indice{1}^\exposant{\text{opt}} \equiv \eta_1^\sensor{ifo} - \tfrac{1}{2} \Theta^\sensor{ldws} + \tfrac{\sqrt{3}}{2} H^\sensor{ldws}
    \nonumber \\
    & N_{z_1}^{\text{sus}} \propto \hat{\phi}_\indice{1}^\exposant{\text{opt}} \equiv \phi_1^\sensor{ifo} + \Phi^\sensor{ldws}
\\[0.2in]
    & N_{x_2}^{\text{sus}} \propto \hat{\theta}_\indice{2}^\exposant{\text{opt}} \equiv \theta_2^\sensor{grs} + \tfrac{\sqrt{3}}{2} \Theta^\sensor{ldws} - \tfrac{1}{2} H^\sensor{ldws}
    \nonumber \\
    & N_{y_2}^{\text{sus}} \propto \hat{\eta}_\indice{2}^\exposant{\text{opt}} \equiv \eta_2^\sensor{ifo} + \tfrac{1}{2} \Theta^\sensor{ldws} + \tfrac{\sqrt{3}}{2} H^\sensor{ldws}
    \nonumber \\
    & N_{z_2}^{\text{sus}} \propto \hat{\phi}_\indice{2}^\exposant{\text{opt}} \equiv \phi_2^\sensor{ifo} + \Phi^\sensor{ldws}.
    \nonumber
\end{empheq}

\subsection{Interpretation and consistency testing}
\label{subsection: Interpretation and consistency testing}

In this section, we present a set of thought experiments which help to interpret physically the error signal combinations derived in the previous section, and demonstrate that they achieve their objectives while ensuring stability and control of the system.

Concentrating first on the longitudinal, suspension scheme, imagine an out-of-loop force signal acting on the spacecraft (e.g. stray thrust, micrometeorite, ...). This excitation must not trigger suspension forces, that is, the suspension coordinates combinations in Equation\,(\myhyperref{eq: scheme_sus_long}) must cancel out. Taking the suspension force applied along the $y$ axis of test mass 1, $F_{y_1}$ as an example, if one denotes the induced displacement due to external of the spacecraft by $\Delta X_{\text{ind}}$ and project it along housing reference frames, one obtains:
\begin{align}
    & F_{y_1}^{\text{cmd}} \propto
    \begin{matrix}
         & -\frac{1}{\sqrt{3}} \left( -\frac{\sqrt{3}}{2} \Delta X_{\text{ind}} \right) \\
         & + \frac{2}{\sqrt{3}} \left( -\frac{\sqrt{3}}{2} \Delta X_{\text{ind}} \right) \\
         & + \left( \frac{1}{2} \Delta X_{\text{ind}} \right)
    \end{matrix}
    = 0,
\end{align}
as required.
Similarly, if one injects a torque excitation around $Z$ on the spacecraft inducing a displacement angle $\Delta \Phi_{\text{ind}}$, the propagation to suspension force along $y$ on test mass 1 coming from levers would be:

\begin{align}
     F_{y_1}^{\text{cmd}} \propto
     & - \frac{1}{\sqrt{3}} \Delta x_\indice{H_2/H_1} \Phi_{\text{ind}}  - \Delta y_\indice{H_2/H_1} \Phi_{\text{ind}} \nonumber \\
     & + \left( \frac{\Delta x}{\sqrt{3}} + \Delta y \right) \Phi_{\text{ind}}
     = 0.
\end{align}
This expression is found by using $\Delta \myvec{r}{H2/H1} = \Delta \myvec{r}{H2/B} - \Delta \myvec{r}{H1/B}$, and expressing the three-lever arm terms as follows:

\begin{multline}
    \Delta x_\indice{\text{H}_1/\text{B}} \Phi_{\text{ind}} \myvec{e}{Y} \cdot \myvec{e}{x_1}  - \Delta y_\indice{\text{H}_1/\text{B}} \Phi_{\text{ind}} \myvec{e}{X} \cdot \myvec{e}{x_1} \\
    = \frac{1}{2} \Delta x_\indice{\text{H}_1/\text{B}} \Phi_{\text{ind}} - \frac{\sqrt{3}}{2} \Delta y_\indice{\text{H}_1/\text{B}} \Phi_{\text{ind}},
\end{multline}

\begin{multline}
    \Delta x_\indice{\text{H}_2/\text{B}} \Phi_{\text{ind}} \myvec{e}{Y} \cdot \myvec{e}{x_2}  - \Delta y_\indice{\text{H}_2/\text{B}} \Phi_{\text{ind}} \myvec{e}{X} \cdot \myvec{e}{x_2} \\
    = -\frac{1}{2} \Delta x_\indice{\text{H}_2/\text{B}} \Phi_{\text{ind}}  - \frac{\sqrt{3}}{2} \Delta y_\indice{\text{H}_2/\text{B}} \Phi_{\text{ind}},
\end{multline}

\begin{multline}
    \Delta x_\indice{\text{H}_1/\text{B}} \Phi_{\text{ind}} \myvec{e}{Y} \cdot \myvec{e}{y_1}  - \Delta y_\indice{\text{H}_1/\text{B}} \Phi_{\text{ind}} \myvec{e}{X} \cdot \myvec{e}{y_1} \\
    = \frac{\sqrt{3}}{2} \Delta x_\indice{\text{H}_1/\text{B}} \Phi_{\text{ind}}  - -\frac{1}{2} \Delta y_\indice{\text{H}_1/\text{B}} \Phi_{\text{ind}}.
\end{multline}

Zero suspension forces along other longitudinal suspension d.o.f can be demonstrated with identical reasoning.

Finally, the same check is done on the new suspension angular coordinate. For instance, exciting the spacecraft attitude angle $\Theta$ with an out-of-loop torque, would trigger a suspension torque around $x_2$, $N_{x_2}^{\text{cmd}}$ of:
\begin{align}
    N_{x_2}^{\text{cmd}} \propto - \left( \tfrac{\sqrt{3}}{2} \Theta_{\text{ind}} \right) + \tfrac{\sqrt{3}}{2} \Theta_{\text{ind}} = 0.
\end{align}
One can also verify that an attitude guidance on the spacecraft will still trigger suspension forces and torques on the test masses, ensuring that they will follow the spacecraft while it orbits the Sun. In this case $\Theta$, $H$ and $\Phi$ shown in Equation\,(\myhyperref{eq: scheme_sus_long}) and (\myhyperref{eq: scheme_sus_angle_tm}) are by definition {\bf error signals} and not sensing outputs: they are inputs of the {\it DFACS}. Injecting guidance signals induces no change on $\Theta$, $H$ and $\Phi$, while producing an apparent change to the measured test mass position and angle relative to the housing. Suspension forces and torques will then be triggered as expected, while S/C jitter around the nominal working point will be rejected by the combinations shown in Equation\,(\myhyperref{eq: scheme_sus_long}) and (\myhyperref{eq: scheme_sus_angle_tm}). This important exception to the isolation is true both for longitudinal and angular suspension of the test masses.

\section{DFACS simulation, demonstration of the control performance and the suspension isolation}
\label{section: simulation}

In this section, we discuss the implementation and testing of the new DFACS scheme with simulations of the closed-loop dynamics of the LISA constellation. At present, there exist three simulation tools in development by the LISA Consortium, with at least one ({\tt LISADyn}) being the object of an upcoming publication: a Linear Time Invariant modelling of the closed-loop system in Matlab, a non-linear capable, Python/C++ dynamics simulation integrated with the LISA End-to-End simulation (LISANode) and a Simulink-SimScape simulator in active development. 
In the remainder of this paper we refer to these simulators as {\tt LISADyn-linear}, {\tt LISADyn} and {\tt LISA-SimScape} respectively. This complete and independent set of tools will afford us extensive and robust testing of this new scheme, in terms of isolation from common-mode disturbances, stability performance and acceleration noise performance. The {\tt LISADyn-linear} will be used preferably for design, while the other two will be utilized for validation and robustness checking.  

\begin{table*}[t]
\centering
\caption{Table of sensing and actuation noise settings in the simulations. The $f_c : \alpha$ columns indicate the corner frequency $f_c$ and the slope $\alpha$ of the noise models.}
\label{table: NoiseLevels}
\begin{tabular}{|c||c|c|c||c|c|c|}
\hline
\# & \bf{Sensing Channel} & \bf{\hspace{1 mm} Noise Floor \hspace{1 mm}} & $f_c : \alpha$ & \bf{Actuation Channel} & \bf{\hspace{1 mm} Noise Floor \hspace{1 mm}} & $f_c : \alpha$ \\ \hline

1 & $x_1^{\text{ifo}} / x_2^{\text{ifo}} \ (\si{\metre\persqrthz})$ & $1.0\times 10^{-12}$ &  $1 \si{\milli\hertz}: \sfrac{1}{f^2}$   & Thrust $X  \ (\si{\newton\persqrthz})$     &                                      $2.2\times 10^{-7}$ & $0.5\si{\milli\hertz}: \sfrac{1}{f}$ \\ \hline

2 & $\eta_1^{\text{ifo}} / \eta_2^{\text{ifo}} \ (\si{\radian\persqrthz})$ & $2.0\times 10^{-9}$ & $0.7\si{\milli\hertz}: \sfrac{1}{f^2}$ & Thrust $Y  \ (\si{\newton\persqrthz})$ & $1.3\times 10^{-7}$ & $0.5\si{\milli\hertz}: \sfrac{1}{f}$ \\ \hline

3 & $\phi_1^{\text{ifo}} / \phi_2^{\text{ifo}} \ (\si{\radian\persqrthz})$ & $2.0\times 10^{-9}$ & $0.7\si{\milli\hertz}: \sfrac{1}{f^2}$ & Thrust $Z  \ (\si{\newton\persqrthz})$ & $3.6\times 10^{-7}$ & $0.5\si{\milli\hertz}: \sfrac{1}{f}$ \\ \hline

4 & $x_1^{\text{grs}} / x_2^{\text{grs}} \ (\si{\metre\persqrthz})$ & $1.8\times 10^{-9}$ & $\hspace{1 mm}1\si{\milli\hertz}: \sfrac{1}{\sqrt{f}}$ &  Thrust $\Theta \ (\si{\newton\metre\persqrthz})$ & $7.7\times 10^{-8}$     &  $0.5\si{\milli\hertz}: \sfrac{1}{f}$ \\ \hline

5 & $y_1^{\text{grs}} / y_2^{\text{grs}} \ (\si{\metre\persqrthz})$ & $1.8\times 10^{-9}$ & $\hspace{1 mm}1\si{\milli\hertz}: \sfrac{1}{\sqrt{f}}$ & Thrust $H\ (\si{\newton\metre\persqrthz})$      & $6.9\times 10^{-8}$ & $0.5\si{\milli\hertz}: \sfrac{1}{f}$ \\ \hline

6 & $z_1^{\text{grs}} / z_2^{\text{grs}} \ (\si{\metre\persqrthz})$ & $3.0\times 10^{-9}$ & $\hspace{1 mm}1\si{\milli\hertz}: \sfrac{1}{\sqrt{f}}$ & Thrust $\Phi\ (\si{\newton\metre\persqrthz})$ & $1.3\times 10^{-7}$ & $0.5\si{\milli\hertz}: \sfrac{1}{f}$ \\ \hline

7 & $\theta_1^{\text{grs}} / \theta_2^{\text{grs}} \ (\si{\radian\persqrthz})$ & $120.0\times 10^{-9}$ & $1\si{\milli\hertz}: \sfrac{1}{\sqrt{f}}$ &  $Fy^{\text{grs}} \ (\si{\newton\persqrthz})$      &  $6.0\times 10^{-15}$ & $1.5\si{\milli\hertz}: \sfrac{1}{f}$ \\ \hline

8 & $\eta_1^\sensor{ldws} / \eta_2^\sensor{ldws} \ (\si{\radian\persqrthz})$ & $0.2\times 10^{-9}$ & $0.7\si{\milli\hertz}: \sfrac{1}{f^2}$ &  $Fz^{\text{grs}} \ (\si{\newton\persqrthz})$      &  $10.0\times 10^{-15}$ &  $1.5\si{\milli\hertz}: \sfrac{1}{f}$ \\ \hline

9 & $\phi_1^\sensor{ldws} / \phi_2^\sensor{ldws} \ (\si{\radian\persqrthz})$ & $0.2\times 10^{-9}$ & $0.7\si{\milli\hertz}: \sfrac{1}{f^2}$ &  $N_{x}^{\text{grs}} \ (\si{\newton\metre\persqrthz})$      &  $1.0\times 10^{-15}$ &  $1.5\si{\milli\hertz}: \sfrac{1}{f}$ \\ \hline

10 &           &            &              &  $N_{y}^{\text{grs}} \ (\si{\newton\metre\persqrthz})$      &  $1.0\times 10^{-15}$ & $1.5\si{\milli\hertz}: \sfrac{1}{f}$ \\ \hline

11 &          &            &              &  $N_{z}^{\text{grs}} \ (\si{\newton\metre\persqrthz})$      & $1.0\times 10^{-15}$ & $1.5\si{\milli\hertz}: \sfrac{1}{f}$ \\ \hline
\end{tabular}
\end{table*}

\subsection{Simulation software and control laws}
\label{subsection: Simulation software and control laws}

{\tt LISADyn} simulations inherit their control laws from the LISA Pathfinder mission, for which a similar linear simulation was developed \cite{weyrich_ssm_2008} and proved to be very useful in the context of LISA Pathfinder data analysis and stability performance studies \cite{lisa_pathfinder_collaboration_lisa_2019}.
This LISA Pathfinder closed-loop simulation used a set of control laws designed and provided by industry, in the form of $15$ transfer functions designed to robustly stabilize the $15$ \gls{siso} systems of the decoupled dynamics problem---that is, a system which has been transformed into the d.o.f. space so that a given actuation channel acts only on a single d.o.f. These control laws can be seen as generic second-order dynamics system stabilizers, with spectral shape adapted to Attitude, Suspension and Drag-Free control according to the mission requirements. Such a set of control laws is not yet available to the LISA Consortium, since the DFACS design is an on-going effort under the responsibility of the industrial contractors bidding to build LISA. Given the importance and the impact of DFACS on the noise budget, however, the LISA Consortium is also carrying out this independent activity to evaluate mission performance and prepare for data analysis efforts.

\begin{figure*}[htb]
    \centering
    \subfloat[Longitudinal jitter ($Z$-axis) \label{figure: LongJitterReq}]
    {\includegraphics[scale=0.34, trim={0.0cm 0.0cm 0.0cm 0.0cm}, clip]{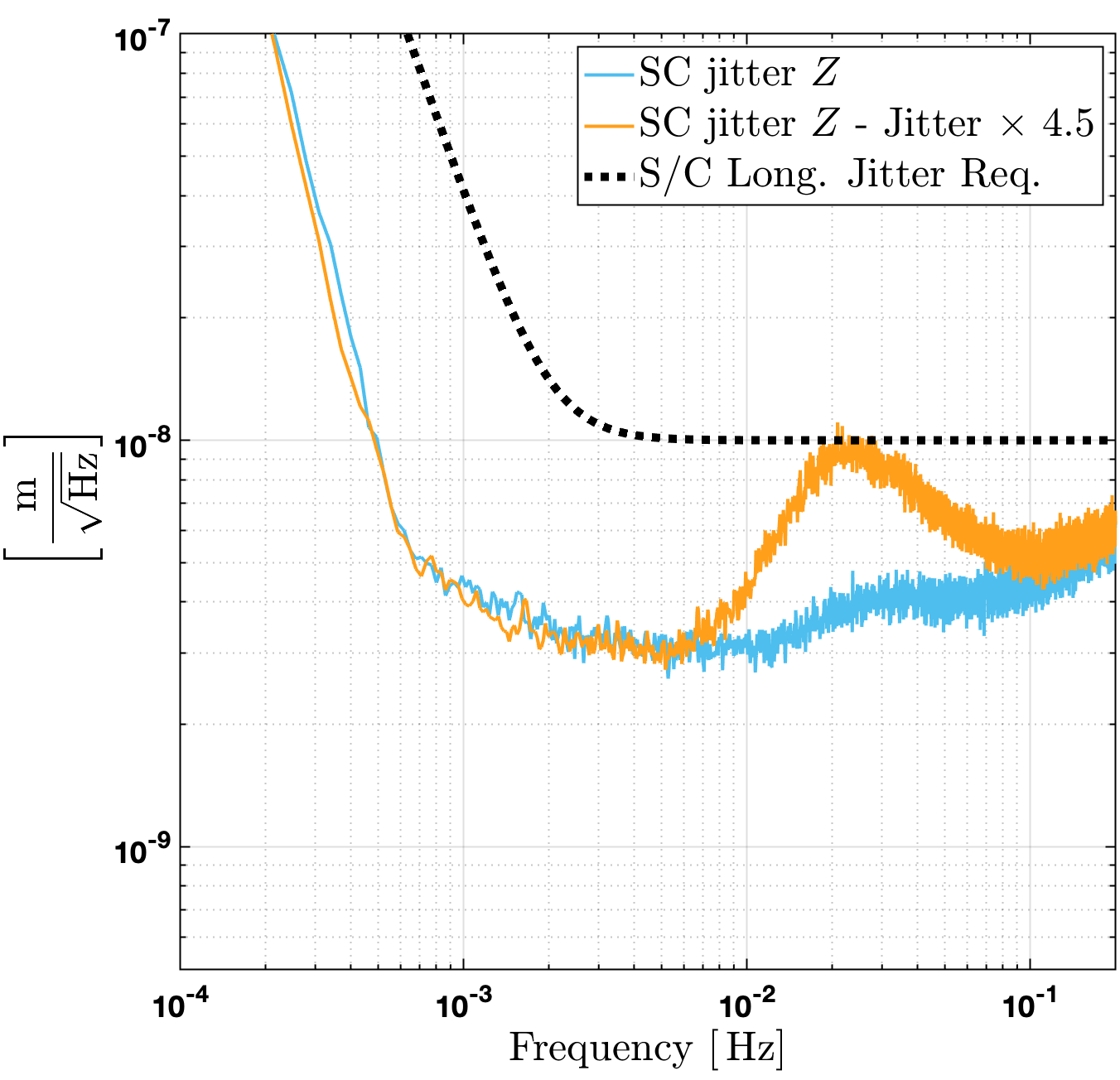}}
    \hspace{1.0cm}
    \subfloat[Angular jitter ($\Phi$ angle) \label{figure: AngJitterReq}]
    {\includegraphics[scale=0.34, trim={0.0cm 0.0cm 0.0cm 0.0cm}, clip]{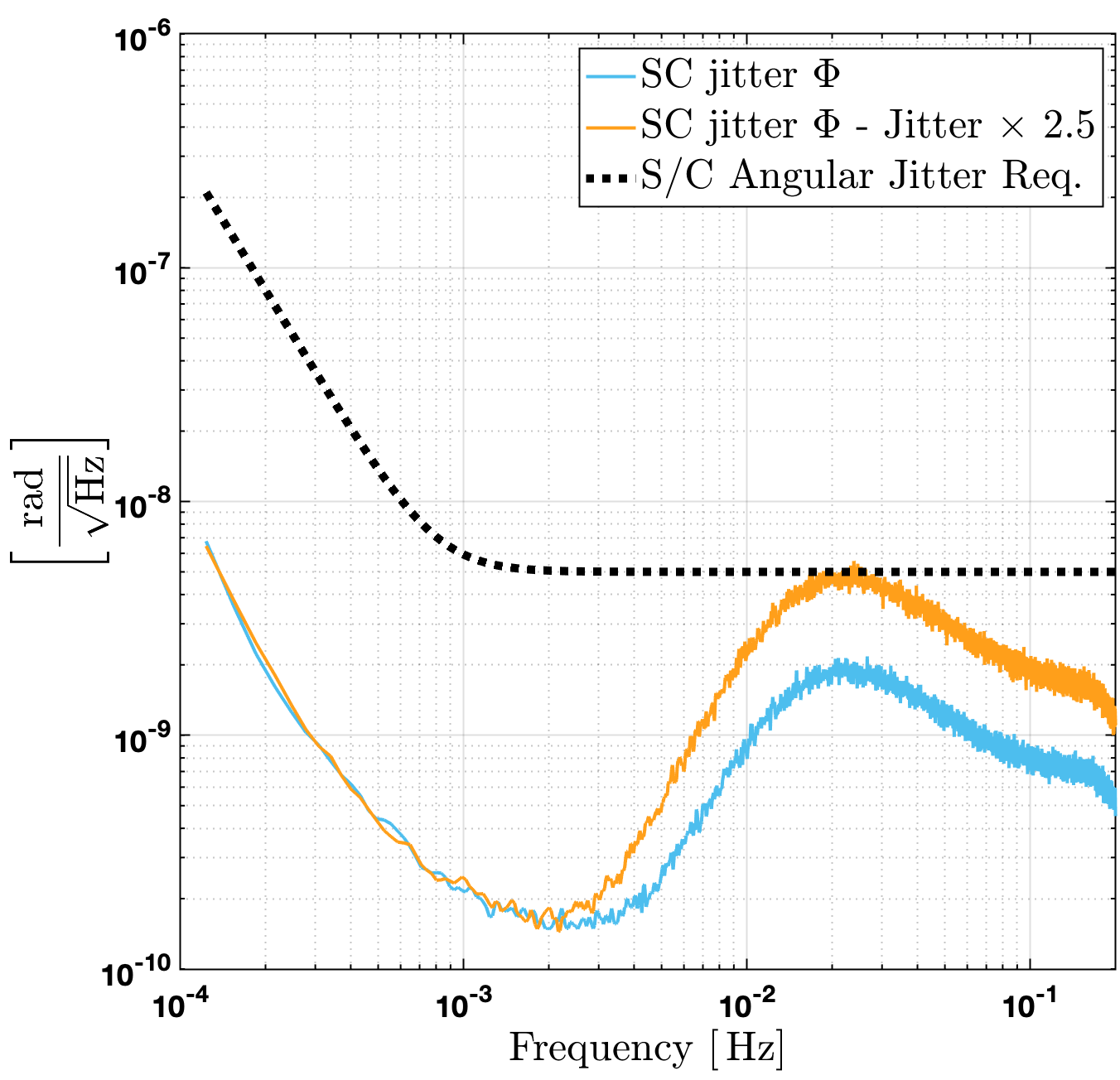}}
    \caption{Achieved stability performance by the simulated DFACS vs. LISA spacecraft (S/C) jitter requirements. $Z$ and $\Phi$ d.o.f are represented as they are leading contributor to actuation cross-talk noise. In light blue are traced the stability curves for nominal thrust noise floor levels (from LISAPathfinder \cite{lisa_pathfinder_collaboration_lisa_2019, armano_lisa_2019-1}). In orange the same curves with increased longitudinal thrust noise ($\times 4.5$) and angular thrust noise ($\times 2.5$) emulating degradation of stability performance down to requirements (at 20 \si{\milli\hertz}). Settings of the orange curve will be used for the suspension scheme testing in order to highlight the benefit of its optimization in the conservative context of performance at the level of requirements.}
\label{figure: JitterReq}
\end{figure*}

A reasonable starting point for the closed-loop system is given by LISA Pathfinder control laws, which can be easily adapted to LISA in applying the coupling matrix which is specific to the LISA spacecraft geometrical and dynamical properties. From the set of independent $15$ \gls{siso} control transfer functions $K^{\text{SISO}}$, one can retrieve the coupled \gls{mimo} control functions $K$ which stabilize the actual LISA dynamics as follows:
\begin{align}
    & K_{\text{att}} = \left[ S_{\text{att}} B_{\text{att}} \right]^{-1} K_{\text{att}}^{\text{SISO}} S_{\text{att}} \\
    & K_{\text{df}} = \left[ S_{\text{df}} B_{\text{df}} \right]^{-1} K_{\text{df}}^{\text{SISO}} S_{\text{df}} \\
    & K_{\text{sus}} = \left[ S_{\text{sus}} B_{\text{sus}} \right]^{-1} K_{\text{sus}}^{\text{SISO}} S_{\text{sus}},
\end{align}
where $B_{\text{att}}$, $B_{\text{df}}$ and $B_{\text{sus}}$ encode and map the contributions of the various input forces and torques applied on the spacecraft and the test masses to the controlled dynamical d.o.f., or simply put, account for the couplings between the actuation channels. $S_{\text{att}}$, $S_{\text{df}}$ and $S_{\text{sus}}$ are the selection matrices which define the sensor channel combination to be used for each control port: the control mapping scheme the DFACS uses. Table \myhyperref{table: control bandwidth} gives the resulting control bandwidth used in the simulator for drag-free, attitude and suspension control, together with the impacted d.o.f, while Table \myhyperref{table: Coupling Matrix} summarizes in a matrix form the drag-free and suspension combinations derived in section \myhyperref{section: DFACS scheme}. Doing so, one obtains a stable closed-loop system with stability performance compatible with LISA requirements---although not yet optimized in term of robustness and model uncertainty. The achieved stability performance turns out to be well below the spacecraft jitter requirements for LISA, both longitudinal and angular. Fig. \myhyperref{figure: JitterReq} shows a comparison between the jitter performance of {\tt LISADyn} \gls{dfacs} and LISA requirement curves.

\begin{table}%[H] add [H] placement to break table across pages
\scriptsize
\caption{\gls{dfacs} control bandwidth for drag-free, attitude and suspension control. Drag-free and attitude control compensates for external disturbances on spacecraft and operates at higher frequency than suspension control, to mitigate low-frequency and D.C. differential drift between the test masses (mostly due by spacecraft self-gravity \cite{armano_constraints_2016}). Above $10 \si{\milli\hertz }$, drag-free and attitude control gain have decreased enough to let force and torque noise through, inducing measurable spacecraft jitter.}
\label{table: control bandwidth}
\begin{ruledtabular}
\begin{tabular}{|c|c|c|}
Control & Degrees of freedom & Bandwidth
\\
\hline
Drag-Free & $x_\indice{1}^\sensor{ifo}$, $x_\indice{2}^\sensor{ifo}$, $z_\indice{1}^\sensor{grs} + z_\indice{2}^\sensor{grs}$ & 200 \si{\milli\hertz}
\\
Attitude & $\Theta^\sensor{ldws}$, $H^\sensor{ldws}$, $\Phi^\sensor{ldws}$ & 200 \si{\milli\hertz}
\\
Suspension (long.) & $y_\indice{1}^\sensor{ifo}$, $y_\indice{2}^\sensor{ifo}$, $z_\indice{1}^\sensor{grs} - z_\indice{2}^\sensor{grs}$ & 1.5 \si{\milli\hertz}
\\
Suspension (rot.) & $\theta_\indice{1}^\sensor{grs}$, $\eta\indice{1}^\sensor{ifo}$, $\phi_\indice{1}^\sensor{ifo}$, $\theta_\indice{2}^\sensor{grs}$, $\eta\indice{2}^\sensor{ifo}$, $\phi_\indice{2}^\sensor{ifo}$ & 1.5 \si{\milli\hertz}
\end{tabular}
\end{ruledtabular}
\end{table}

The stability performance are certainly not guaranteed to reach the level of the simulation, as numerous criteria remain to be implemented in their design (for example robustness w.r.t. model uncertainties, constraints from accurate modeling of sensors and actuators). At this early stage, therefore, we choose a conservative approach to illustrate the benefit of the newly proposed suspension scheme and assume a degraded level of stability, such that the performance coincides with the requirement limits.
We mimic a degraded spacecraft stability by increasing the noisy forces and torques applied on the spacecraft. We find that multiplying the forces and torques by factors $4.5$ and $2.5$ respectively compared levels observed with LISA Pathfinder \cite{lisa_pathfinder_collaboration_lisa_2019, armano_lisa_2019-1} the simulated stability spectra are brought up to the requirement threshold, as shown by the light blue and orange traces of Figure\,\myhyperref{figure: JitterReq}.

In the following sections, we present simulation results using {\tt LISADyn-linear} as it provides better flexibility for investigation and testing. The degraded stability case will be considered to reflect the LISA requirement levels and highlights the benefit of optimizing the suspension scheme. The results and conclusions have been cross-checked and validated by more comprehensive simulation implementations such as {\tt LISADyn} or {\tt LISA-SimScape}.
It is finally worth noting that the development of the proposed suspension scheme has assumed a fixed \gls{mosa} opening angle of $\phi=60 \si{\degree}$ and the simulations discussed below use such approximation. 
In reality, $\phi$ will vary by up to $\pm 1 \si{degree}$ in order to account for variations in the spacecraft orbits.
Performing a first order Taylor expansion of the formulae \myhyperref{eq: scheme_sus_long} and \myhyperref{eq: scheme_sus_angle_tm} around $\phi = 60 \si{\degree}$ and running the same simulations including \gls{mosa} rotations shows that the constant opening approximation leads to a $\approx 2 \%$ residual level in the suspension isolation at most
The approximation therefore guarantees a $98 \%$ rejection of platform jitter from suspension control while allowing us to keep a simpler, time-invariant control scheme.

\begin{table*}[t]
\centering
\caption{DFACS optimized scheme represented through the mapping matrix from the sensing error signals to the commanded forces / torques. The symbol "-" replaces the $0.0$ value to enhance readability of the table.}
\label{table: Coupling Matrix}
\begin{tabular}{cc||c|c|c|c|c|c|c|c|c|c|c|c|c|c|c|c|c|c|c|c|c|}
\multicolumn{2}{c||}{Control} & $\Theta^\sensor{ldws}$ & $H^\sensor{ldws}$ & $\Phi^\sensor{ldws}$ & $x_\indice{1}^\sensor{ifo}$ & $\eta_\indice{1}^\sensor{ifo}$ & $\phi_\indice{1}^\sensor{ifo}$ & $x_\indice{2}^\sensor{ifo}$ & $\eta_\indice{2}^\sensor{ifo}$ & $\phi_\indice{2}^\sensor{ifo}$ & $x_\indice{1}^\sensor{grs}$ & $y_\indice{1}^\sensor{grs}$ & $z_\indice{1}^\sensor{grs}$ & $\theta_\indice{1}^\sensor{grs}$ & $\eta_\indice{1}^\sensor{grs}$ & $\phi_\indice{1}^\sensor{grs}$ & $x_\indice{2}^\sensor{grs}$ & $y_\indice{2}^\sensor{grs}$ & $z_\indice{2}^\sensor{grs}$ & $\theta_\indice{2}^\sensor{grs}$ & $\eta_\indice{2}^\sensor{grs}$ & $\phi_\indice{2}^\sensor{grs}$ \\
\hline
\noalign{\vskip 2mm}
\hline
\multirow{3}*{Att.}
    & $N_X$ & $1$ & {-} & {-} & {-} & {-} & {-} & {-} & {-} & {-} & {-} & {-} & {-} & {-} & {-} & {-} & {-} & {-} & {-} & {-} & {-} & {-} \\
    \cline{2-23}
    & $N_Y$ & {-} & $1$ & {-} & {-} & {-} & {-} & {-} & {-} & {-} & {-} & {-} & {-} & {-} & {-} & {-} & {-} & {-} & {-} & {-} & {-} & {-} \\
    \cline{2-23}
    & $N_Z$ & {-} & {-} & $1$ & {-} & {-} & {-} & {-} & {-} & {-} & {-} & {-} & {-} & {-} & {-} & {-} & {-} & {-} & {-} & {-} & {-} & {-} \\
\hline
\noalign{\vskip 8mm}
\hline
\multirow{3}*{DF}
    & $F_X$ & {-} & {-} & {-} & $\tfrac{1}{\sqrt{3}}$ & {-} & {-} & $\tfrac{1}{\sqrt{3}}$ & {-} & {-} & {-} & {-} & {-} & {-} & {-} & {-} & {-} & {-} & {-} & {-} & {-} & {-} \\
    \cline{2-23}
    & $F_Y$ & {-} & {-} & {-} & $1$ & {-} & {-} & $-1$ & {-} & {-} & {-} & {-} & {-} & {-} & {-} & {-} & {-} & {-} & {-} & {-} & {-} & {-} \\
    \cline{2-23}
    & $F_Z$ & {-} & {-} & {-} & {-} & {-} & {-} & {-} & {-} & {-} & {-} & {-} & $\tfrac{1}{2}$ & {-} & {-} & {-} & {-} & {-} & $\tfrac{1}{2}$ & {-} & {-} & {-} \\
\hline
\noalign{\vskip 8mm}
\hline
    & $F_{x_\indice{1}}$ & {-} & {-} & {-} & {-} & {-} & {-} & {-} & {-} & {-} & {-} & {-} & {-} & {-} & {-} & {-} & {-} & {-} & {-} & {-} & {-} & {-} \\
    \cline{2-23}
\multirow{3}*{Sus.}
    & $F_{y_\indice{1}}$ & $\tfrac{\Delta z}{\sqrt{3}}$ & $\Delta z$ & $-\tfrac{\Delta x}{\sqrt{3}} - \Delta y$ & -$\tfrac{1}{\sqrt{3}}$ & {-} & {-} & $\tfrac{2}{\sqrt{3}}$ & {-} & {-} & {-} & $1$ & {-} & {-} & {-} & {-} & {-} & {-} & {-} & {-} & {-} & {-} \\
    \cline{2-23}
\multirow{3}*{TM1}
    & $F_{z_\indice{1}}$ & -$\Delta y$ & $\Delta x$ & {-} & {-} & {-} & {-} & {-} & {-} & {-} & {-} & {-} & $1$ & {-} & {-} & {-} & {-} & {-} & $-1$ & {-} & {-} & {-} \\
    \cline{2-23}
    & $N_{x_\indice{1}}$ & $\tfrac{\sqrt{3}}{2}$ & $\tfrac{1}{2}$ & {-} & {-} & {-} & {-} & {-} & {-} & {-} & {-} & {-} & {-} & $1$ & {-} & {-} & {-} & {-} & {-} & {-} & {-} & {-} \\
    \cline{2-23}
    & $N_{y_\indice{1}}$ & -$\tfrac{1}{2}$ & $\tfrac{\sqrt{3}}{2}$ & {-} & {-} & {-} & {-} & {-} & {-} & {-} & {-} & {-} & {-} & {-} & $1$ & {-} & {-} & {-} & {-} & {-} & {-} & {-} \\
    \cline{2-23}
    & $N_{z_\indice{1}}$ & {-} & {-} & $1$ & {-} & {-} & {-} & {-} & {-} & {-} & {-} & {-} & {-} & {-} & {-} & $1$ & {-} & {-} & {-} & {-} & {-} & {-} \\
\hline
\noalign{\vskip 4mm}
\hline
    & $F_{x_\indice{2}}$ & {-} & {-} & {-} & {-} & {-} & {-} & {-} & {-} & {-} & {-} & {-} & {-} & {-} & {-} & {-} & {-} & {-} & {-} & {-} & {-} & {-} \\
    \cline{2-23}
\multirow{3}*{Sus.}
    & $F_{y_\indice{2}}$ & -$\tfrac{\Delta z}{\sqrt{3}}$ & $\Delta z$ & $\tfrac{\Delta x}{\sqrt{3}} - \Delta y$ & -$\tfrac{2}{\sqrt{3}}$ & {-} & {-} & $\tfrac{1}{\sqrt{3}}$ & {-} & {-} & {-} & {-} & {-} & {-} & {-} & {-} & {-} & $1$ & {-} & {-} & {-} & {-} \\
    \cline{2-23}
\multirow{3}*{TM2}
    & $F_{z_\indice{2}}$ & $\Delta y$ & $-\Delta x$ & {-} & {-} & {-} & {-} & {-} & {-} & {-} & {-} & {-} & $-1$ & {-} & {-} & {-} & {-} & {-} & $1$ & {-} & {-} & {-} \\
    \cline{2-23}
    & $N_{x_\indice{2}}$ & $\tfrac{\sqrt{3}}{2}$ & $-\tfrac{1}{2}$ & {-} & {-} & {-} & {-} & {-} & {-} & {-} & {-} & {-} & {-} & {-} & {-} & {-} & {-} & {-} & {-} & $1$ & {-} & {-} \\
    \cline{2-23}
    & $N_{y_\indice{2}}$ & $\tfrac{1}{2}$ & $\tfrac{\sqrt{3}}{2}$ & {-} & {-} & {-} & {-} & {-} & {-} & {-} & {-} & {-} & {-} & {-} & {-} & {-} & {-} & {-} & {-} & {-} & $1$ & {-} \\
    \cline{2-23}
    & $N_{z_\indice{2}}$ & {-} & {-} & $1$ & {-} & {-} & {-} & {-} & {-} & {-} & {-} & {-} & {-} & {-} & {-} & {-} & {-} & {-} & {-} & {-} & {-} & $1$ \\
\hline
\end{tabular}
\end{table*}

\subsection{Experiment: Thrust forces along X, Y and Z}

We first demonstrate the isolation of suspension forces from translational spacecraft jitter. Thrust forces are injected on the spacecraft along $X$, $Y$ and $Z$ d.o.f (along the axes of the spacecraft body frame $\mathcal{B}$-frame), and we verify that these signals do not produce coherent suspension forces. 
%, since such perturbation can and must be corrected through - i.e. by actuating on the spacecraft only. 
We simulate the three following cases:

\begin{itemize}
    \item Simple suspension control scheme where we suspend the test masses locking on coordinates $[ \hat{y}_\indice{1}^\exposant{\text{sim}}, \allowbreak  \hat{y}_\indice{2}^\exposant{\text{sim}}, \allowbreak  \hat{z}_\indice{2}^\exposant{\text{sim}}, \allowbreak  \hat{\theta}_\indice{1}^\exposant{\text{sim}}, \allowbreak  \hat{\eta}_\indice{1}^\exposant{\text{sim}}, \allowbreak  \hat{\phi}_\indice{1}^\exposant{\text{sim}}, \allowbreak  \hat{\theta}_\indice{2}^\exposant{\text{sim}}, \allowbreak  \hat{\eta}_\indice{2}^\exposant{\text{sim}}, \allowbreak  \hat{\phi}_\indice{2}^\exposant{\text{sim}}]$, which are aliases of the d.o.f as they are measured on-board (see appendix \myhyperref{appendix: simple scheme}).
    \item Suspension scheme using the optimized suspension coordinates $[ \hat{y}_\indice{1}^\exposant{\text{opt}}, \allowbreak  \hat{y}_\indice{2}^\exposant{\text{opt}}, \allowbreak  \hat{z}_\indice{2}^\exposant{\text{opt}}, \allowbreak  \hat{\theta}_\indice{1}^\exposant{\text{opt}}, \allowbreak  \hat{\eta}_\indice{1}^\exposant{\text{opt}}, \allowbreak  \hat{\phi}_\indice{1}^\exposant{\text{opt}}, \allowbreak  \hat{\theta}_\indice{2}^\exposant{\text{opt}}, \allowbreak  \hat{\eta}_\indice{2}^\exposant{\text{opt}}, \allowbreak  \hat{\phi}_\indice{2}^\exposant{\text{opt}}]$ as defined in Section\,\myhyperref{section: DFACS scheme} including the lever arm correction (Eq. \myhyperref{eq: scheme_sus_long}) and (Eq. \myhyperref{eq: scheme_sus_angle_tm}).
    \item Suspension scheme using the optimized suspension coordinates as defined in Section\,\myhyperref{section: DFACS scheme} without the lever arm corrections.
\end{itemize}

All dynamical couplings between common and differential modes such as stiffness or sensing and actuation cross-talks have been turned off for these simulations. While they would trigger legitimate suspension forces or torques, their inclusion would complicate the interpretation of the injection experiments making it difficult to assess the achieved level of suspension isolation.

\begin{figure*}[htb]
    \centering
    \subfloat[Suspension forces along y1 \label{figure: LongJitter/Fy1}]
    {\includegraphics[scale=0.27, trim={0.0cm 0.0cm 0.0cm 0.0cm}, clip]{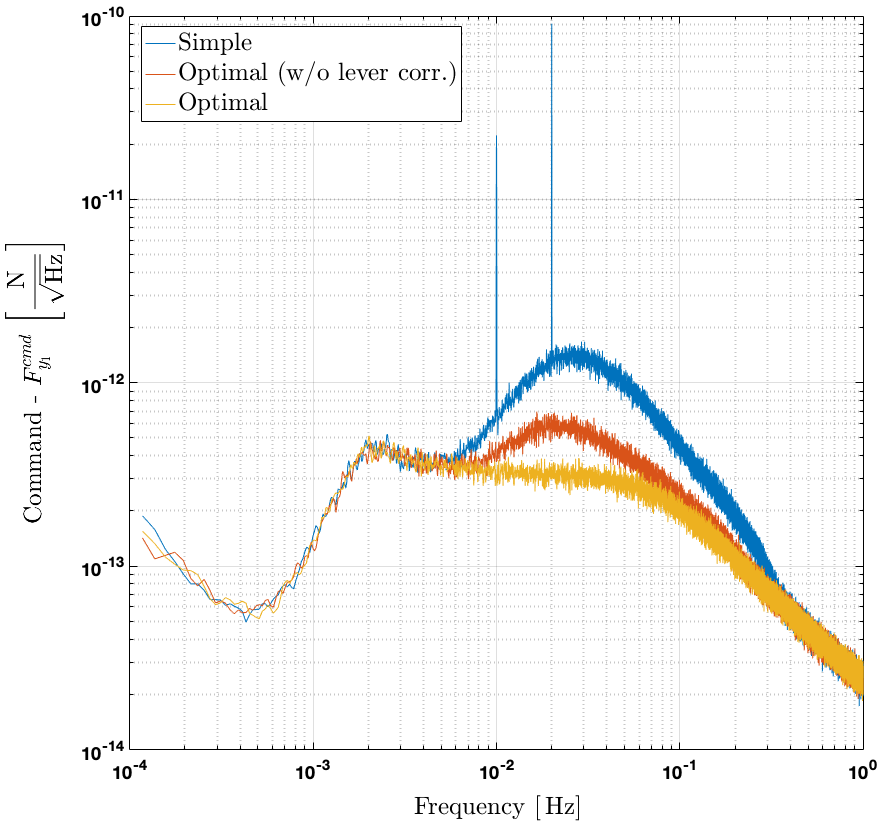}}
    \hspace{0.5cm}
    \subfloat[Suspension forces along z2 \label{figure: LongJitter/Fz2}]
    {\includegraphics[scale=0.27, trim={0.0cm 0.0cm 0.0cm 0.0cm}, clip]{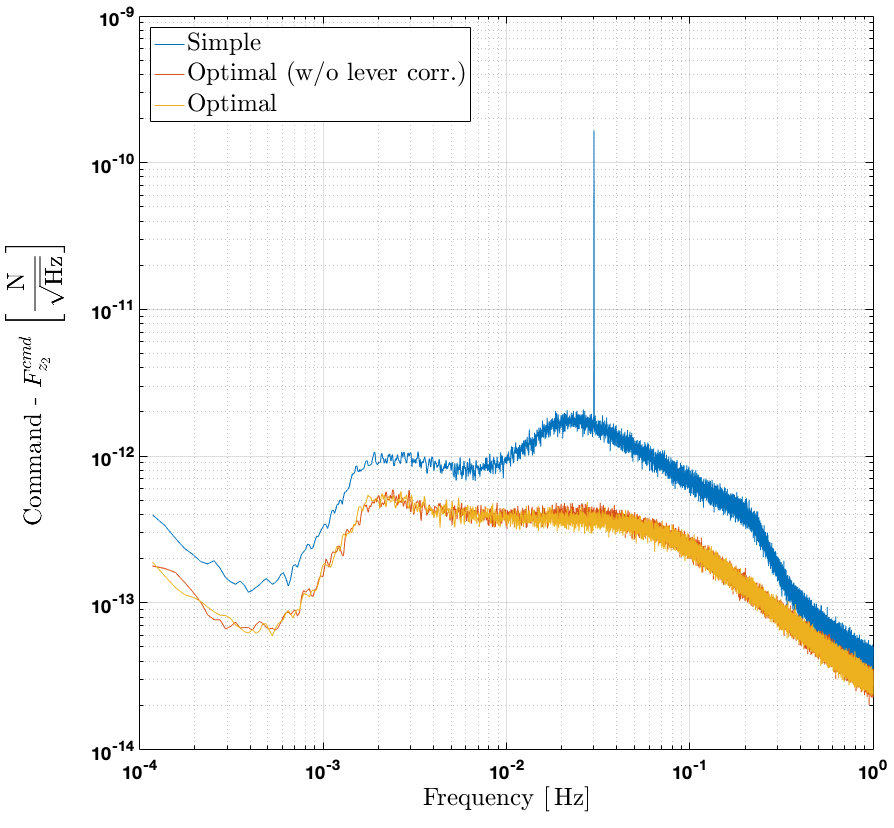}}
    \caption{Suspension response to longitudinal jitter injection on spacecraft (10 \si{\milli\hertz}, 20 \si{\milli\hertz} and 30 \si{\milli\hertz})}
\label{figure: LongJitter}
\end{figure*}

We have injected 1\,\si{\micro\newton} sinusoidal force along $X$, $Y$ and $Z$ simultaneously, with frequencies of 10\,\si{\milli\hertz}, 20\,\si{\milli\hertz} and 30\,\si{\milli\hertz} respectively. Fig. \myhyperref{figure: LongJitter/Fy1} and \myhyperref{figure: LongJitter/Fz2} show the amplitude spectral density (ASD) of the resulting commanded suspension force on test mass 1 along the $y$ axis of its housing ($\myvec{e}{y, H_{\indice{1}}}$), and the suspension forces on test mass 2 along $\myvec{e}{z, H_{\indice{2}}}$. This choice is motivated by the symmetric results observed for $y_1$ and $y_2$, and since only $z_2$ is suspended in the {\it simple} suspension scheme ($z_1$ is drag-free controlled). The figures show clearly that the new schemes (red and yellow traces) null the requested suspension forces below the level of the force noise. One also sees that, besides the injection, the overall force level in the considered bandwidth is significantly decreased demonstrating the isolation from spacecraft motion that is the aim of this work.
Accounting for lever-arm effects (yellow trace) produces improved performance because suspension forces are more effectively isolated from spacecraft rotational jitter in this case. This will be investigated further in the next section, where angular injection on the spacecraft will be tested.

\subsection{Experiment: Thrust torques around X, Y and Z}

The case of test mass suspension force isolation from spacecraft rotation is somewhat more complex than the translational case because of the various inertial and lever arms effects projecting along the housing frame axes. Torque injection experiments provide a comprehensive test of the jitter isolation scheme, and from the three considered test cases, a clear demonstration of the difference and advantages of the derived schemes.

In this set of experiments 1\,\si{\micro\newton \metre} sinusoidal torques around $X$, $Y$ and $Z$ have been injected simultaneously, at 10\,\si{\milli\hertz}, 20\,\si{\milli\hertz} and 30\,\si{\milli\hertz} respectively. We present ASDs of the suspension forces and torques: $F_{y_\indice{1}}$, $F_{z_\indice{2}}$, $N_{y_\indice{2}}$, $N_{z_\indice{1}}$, as examples in order to show instances of each suspension d.o.f.
No significant difference between test masses 1 and 2 has been observed.

\begin{figure*}[htb]
    \centering
    \subfloat[Suspension forces along y1 \label{figure: AngJitter/Fy1}]
    {\includegraphics[scale=0.33, trim={0.0cm 0.0cm 0.0cm 0.0cm}, clip]{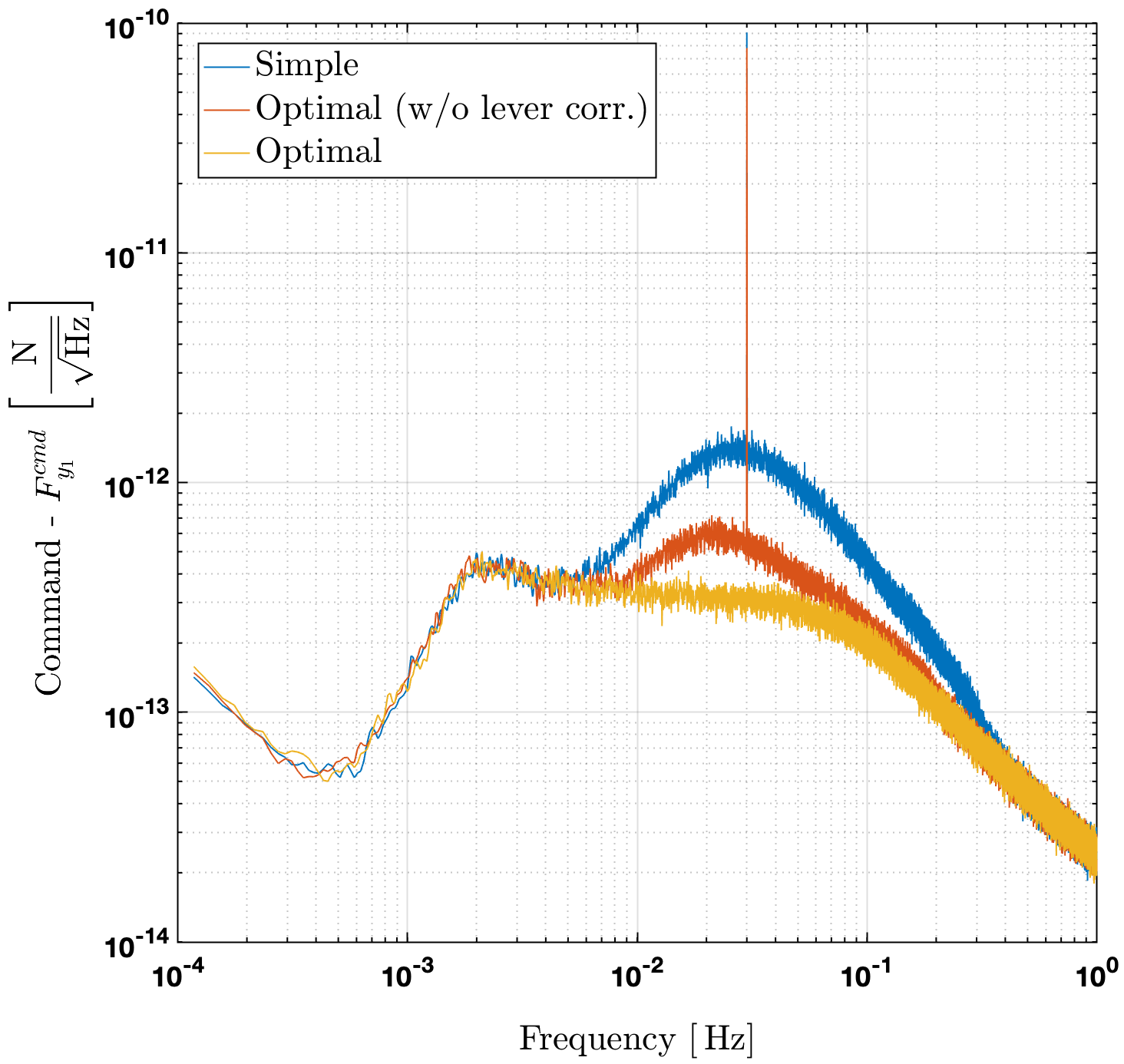}}
    \hspace{0.5cm}
    \subfloat[Suspension forces along z2 \label{figure: AngJitter/Fz2}]
    {\includegraphics[scale=0.33, trim={0.0cm 0.0cm 0.0cm 0.0cm}, clip]{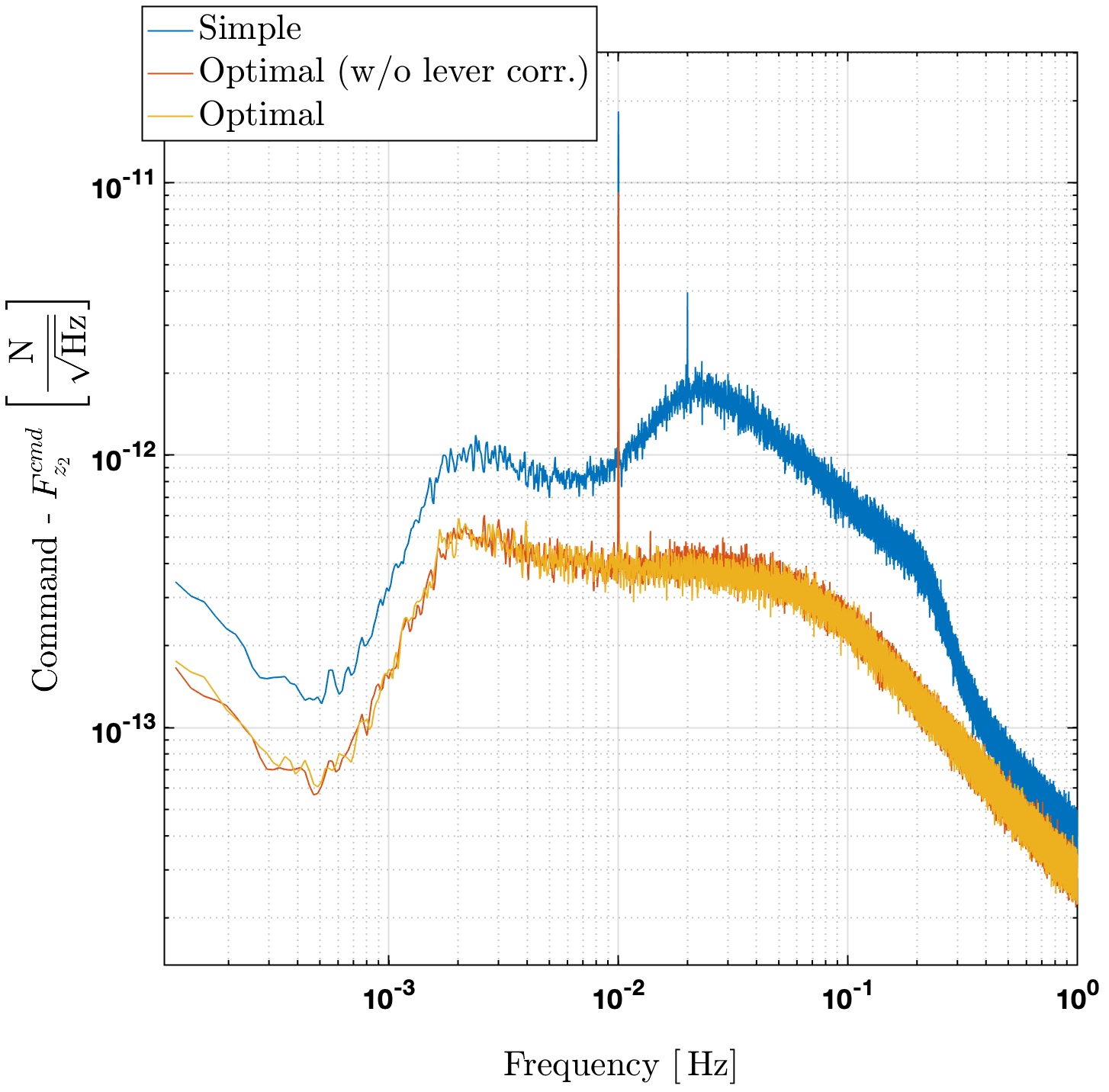}}
    \\
    \subfloat[Suspension torques around y2 \label{figure: AngJitter/Ty2}]
    {\includegraphics[scale=0.33, trim={0.0cm 0.0cm 0.0cm 0.4cm}, clip]{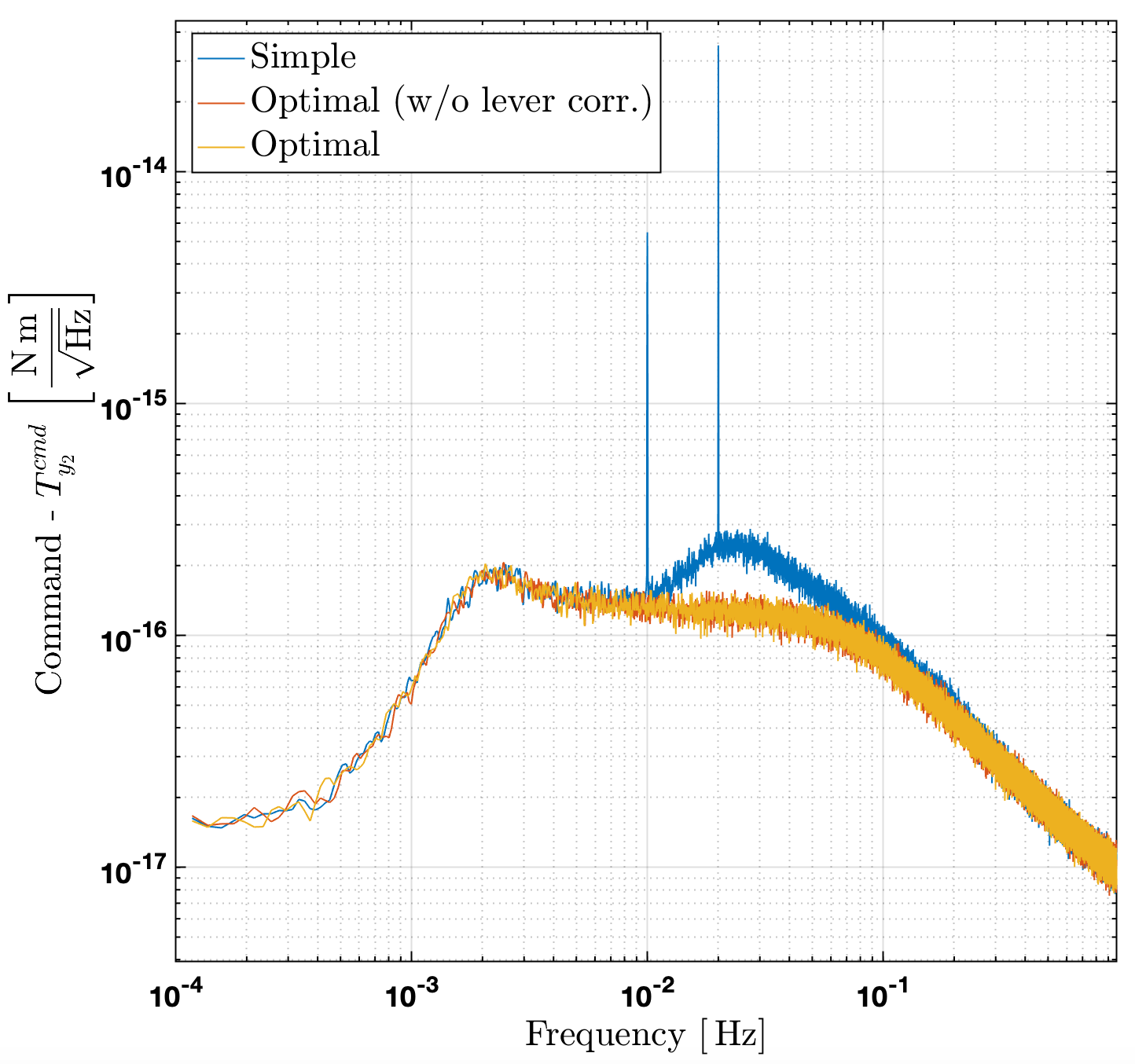}}
    \hspace{0.5cm}
    \subfloat[Suspension torques around z1 \label{figure: AngJitter/Tz1}]
    {\includegraphics[scale=0.33, trim={0.0cm 0.0cm 0.0cm 0.0cm}, clip]{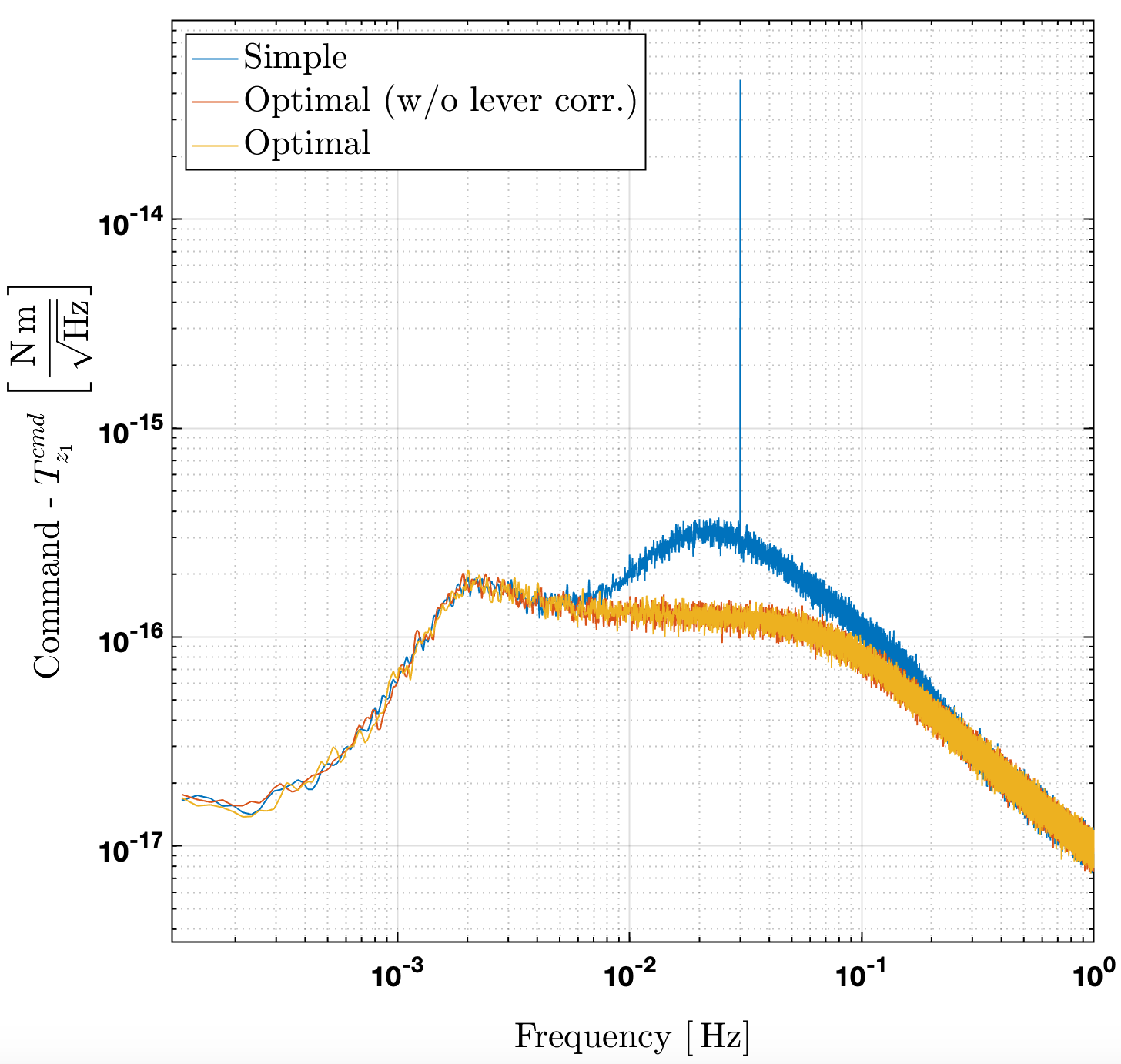}}
    \caption{Suspension response to angular jitter injection on spacecraft (10 \si{\milli\hertz}, 20 \si{\milli\hertz} and 30 \si{\milli\hertz})}
\label{figure: AngJitter}
\end{figure*}

Focusing first on the longitudinal requested suspension forces $F_{y_\indice{1}}$ and $F_{z_\indice{2}}$ shown in Fig.\,\myhyperref{figure: AngJitter/Fy1} and Fig.\,\myhyperref{figure: AngJitter/Fz2}, we confirm that the full isolation scheme compensating for the levers efficiently suppresses the three injections to a level below the noise. We also see that the scheme without lever correction (red traces) leaves large peaks due to coupling to the rotational injection through lever arms, demonstrating again the necessity of this additional correction.

The experiment also demonstrates isolation from spacecraft jitter of the test mass angular suspension. Using the updated angular suspension coordinates as described in Equation\,(\myhyperref{eq: scheme_sus_angle_tm}), one efficiently suppresses the three injection peaks in the suspension torques as seen in Figures \myhyperref{figure: AngJitter/Ty2} and \myhyperref{figure: AngJitter/Tz1} showing commanded torque spectra, $N_\indice{y_2}$ and $N_\indice{z_1}$ respectively. One also notes a significant decrease of the noise floor around 20\,\si{\milli\hertz} for $N_\indice{y_2}$ and $N_\indice{z_1}$  where suspension torque noise is driven by motion rather than sensor noise.

\subsection{Experiment: angular guidance of the spacecraft}

\begin{figure*}[htb]
    \centering
    \subfloat[Suspension forces along $y_1$ \label{figure: AttGuidance/Fy1}]
    {\includegraphics[scale=0.33, trim={0.0cm 0.0cm 0.0cm 0.0cm}, clip]{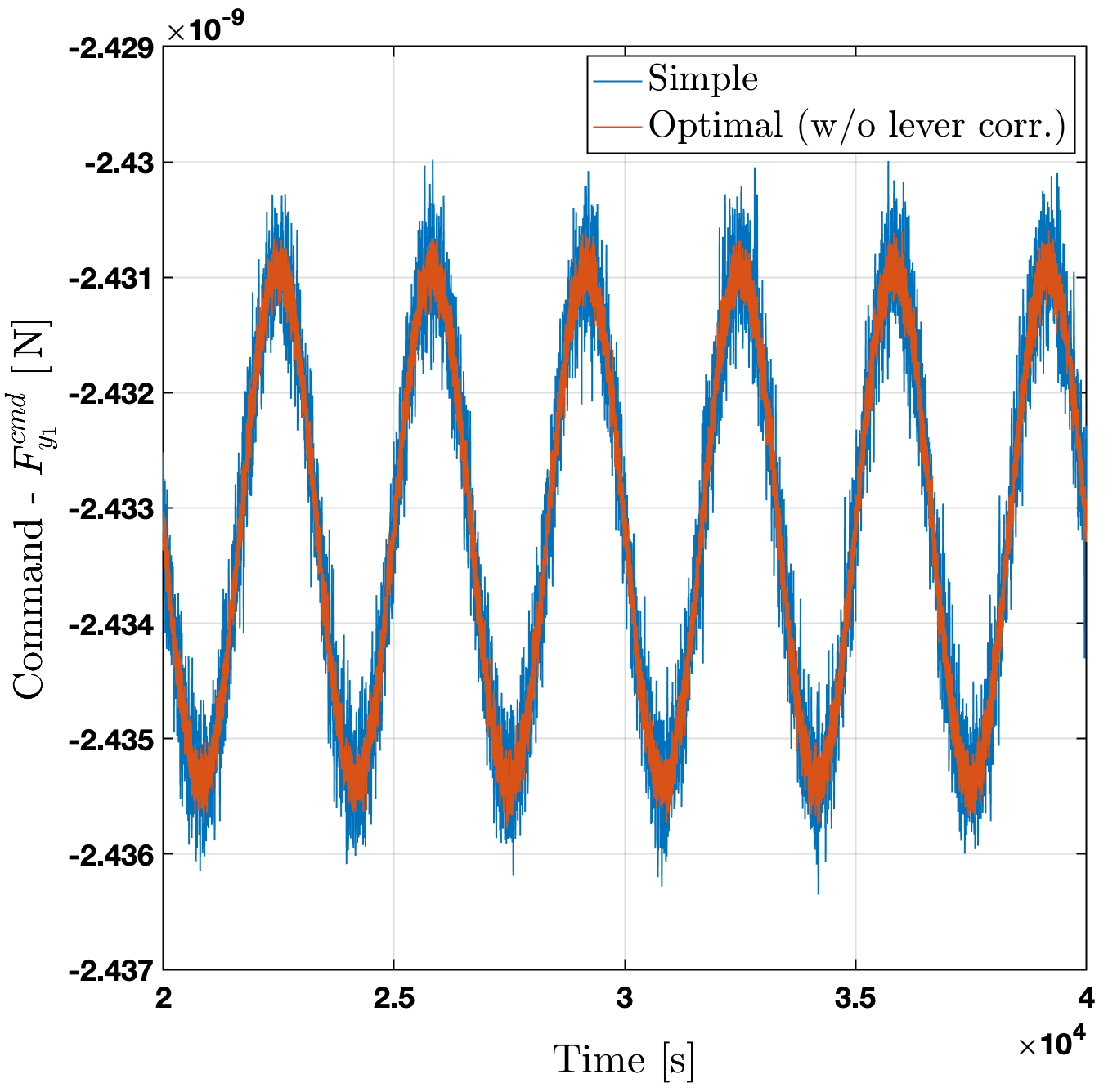}}
    \hspace{0.5cm}
    \subfloat[Suspension torques around $z_2$ \label{figure: AttGuidance/Tz2}]
    {\includegraphics[scale=0.33, trim={0.0cm 0.0cm 0.0cm 0.0cm}, clip]{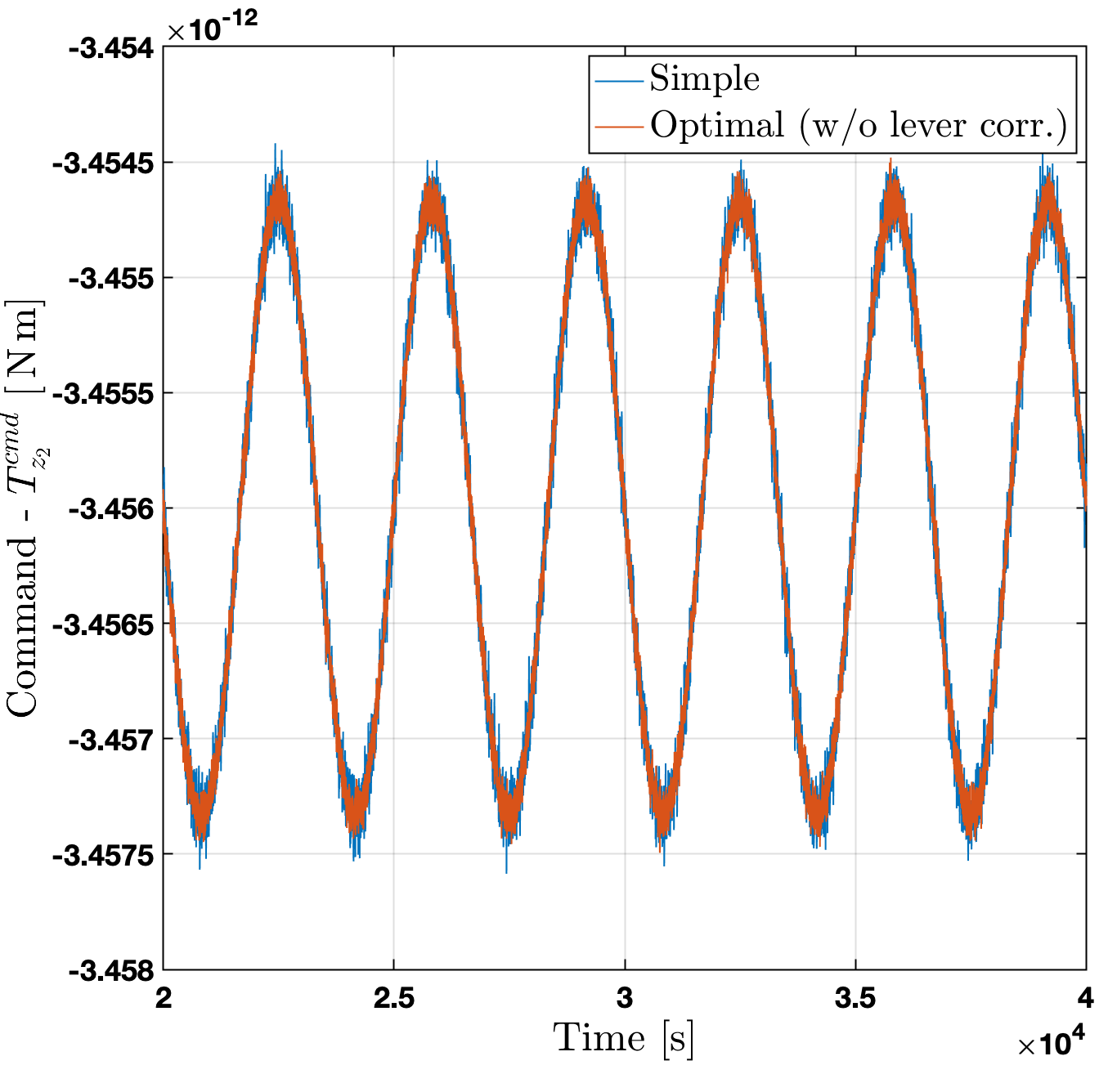}}
    \caption{Time series suspension response to attitude guidance injection on spacecraft at 0.1 \si{\milli\hertz}, 0.2 \si{\milli\hertz} and 0.3 \si{\milli\hertz} around $X$, $Y$ and $Z$ respectively.}
\label{figure: AttGuidance}
\end{figure*}

It is important to ensure, as briefly discussed in Section \myhyperref{subsection: Interpretation and consistency testing}, that the test masses follow the spacecraft as they rotate to maintain laser links with each other. In practice, suspension will have to push the test masses in order to go along with the constellation rotation, as well as correct the test mass orientation in their housings accordingly, both compensations act at low frequency, out of the measurement band. One must then verify that the isolation from common-mode jitters does not suppress this low-frequency rigidity. As explained in \myhyperref{subsection: Interpretation and consistency testing}, this can be tested by verifying that attitude guidance signals trigger corresponding suspension forces and torques on test masses. We perform a simulated experiment where guidance spacecraft attitude signals of 1\,\si{\nano\radian} around $X$, $Y$ and $Z$ are injected at 0.1\,\si{\milli\hertz}, 0.2\,\si{\milli\hertz} and 0.3\,\si{\milli\hertz} respectively. Fig.\,\myhyperref{figure: AttGuidance} shows an extract from the time series of the suspension force along $y_1$ and torque around $z_2$. The clear oscillation at 0.3\,\si{\milli\hertz} confirms the response of suspension to attitude guidance---and hence to constellation rotation---is intact. Similar responses are seen at 0.1\,\si{\milli\hertz} and 0.2\,\si{\milli\hertz} on other d.o.f. not shown here.% but analyzed for completeness.

\subsection{Control performance and residual jitter levels}

Finally, one must check that the new suspension scheme does not increase spacecraft-to-test masses jitter for the corresponding suspended d.o.f., that would result in an increased acceleration noise through stiffness couplings. 
Intuitively, one may imagine that prohibiting direct actuation on the test masses to compensate for spacecraft jitter would worsen position stability between spacecraft and test masses. We show below that it is not the case, since in a simple scheme where $y$ and $z$ test mass coordinates are suspended, suspension actuation competes with drag-free control to achieve the same result.

\begin{figure*}[htb]
    \centering
    \subfloat[TM1-to-S/C jitter along $y_1$ \label{figure: AngJitter/y1}]
    {\includegraphics[scale=0.23, trim={0.0cm 0.0cm 0.0cm 0.0cm}, clip]{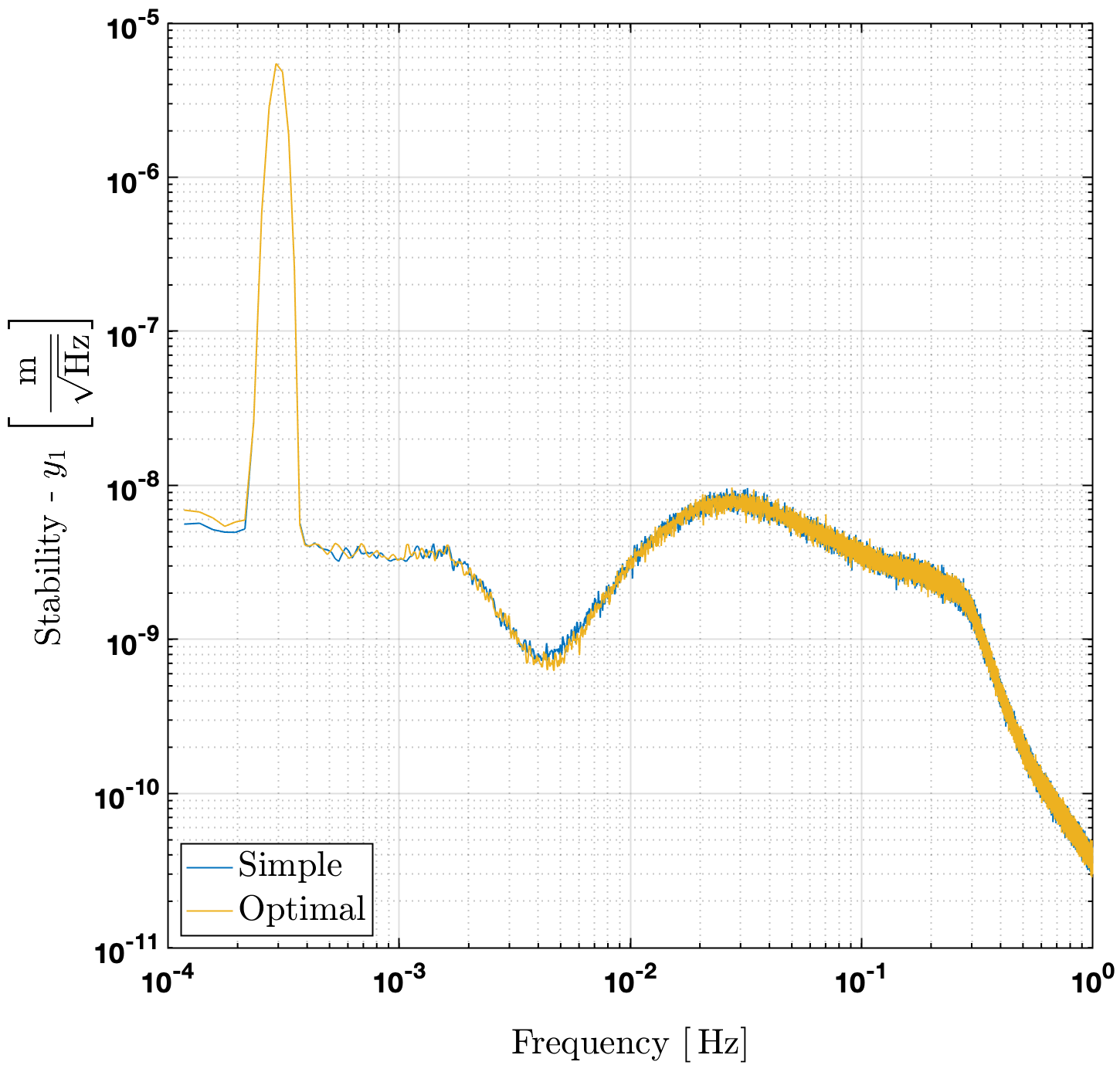}}
    \subfloat[TM2-to-S/C jitter around $x_2$ \label{figure: AngJitter/theta2}]
    {\includegraphics[scale=0.23, trim={0.0cm 0.0cm 0.0cm 0.0cm}, clip]{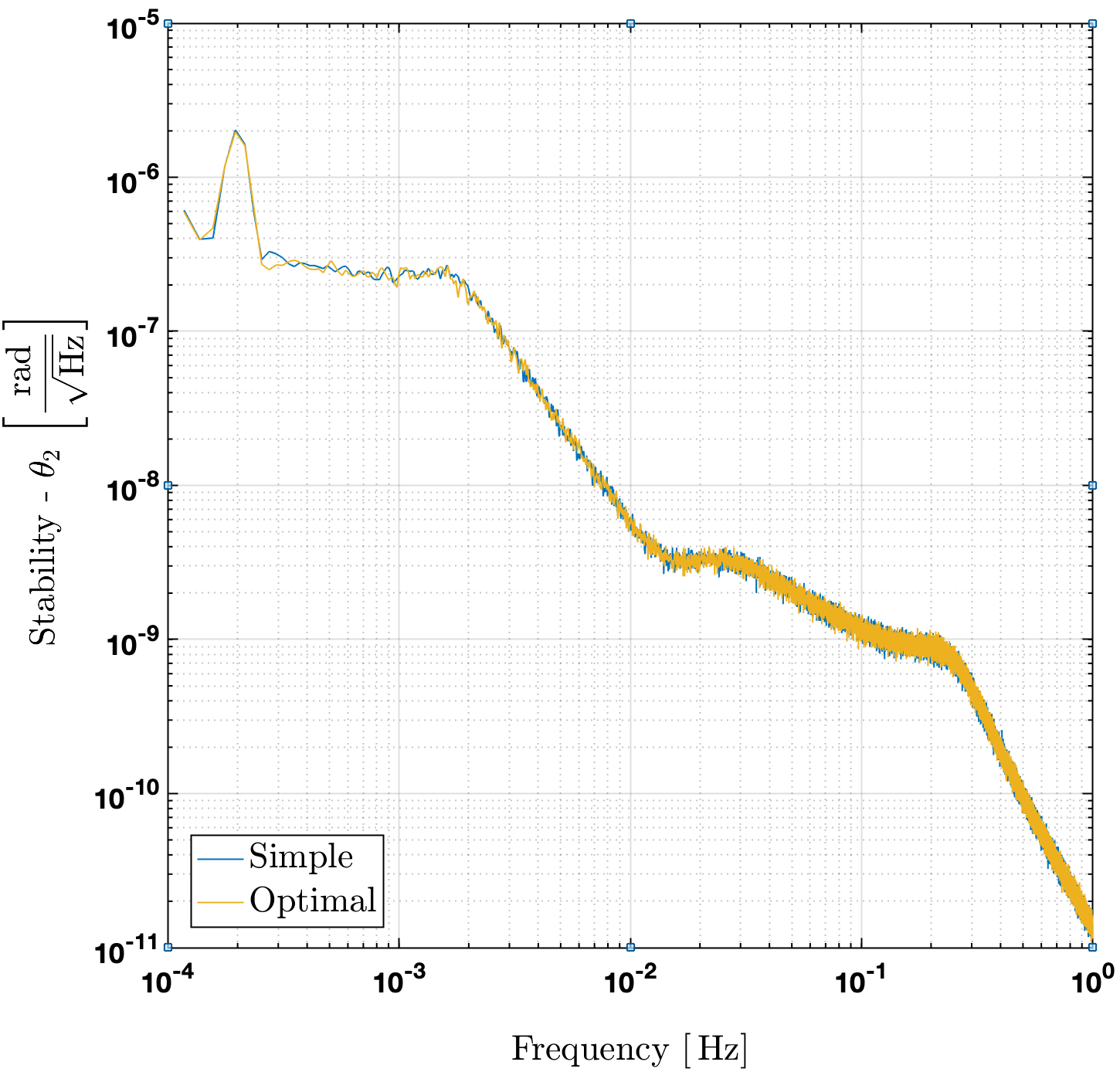}}
    \subfloat[TM1-to-S/C jitter around $z_1$ \label{figure: AngJitter/phi1}]
    {\includegraphics[scale=0.23, trim={0.0cm 0.0cm 0.0cm 0.0cm}, clip]{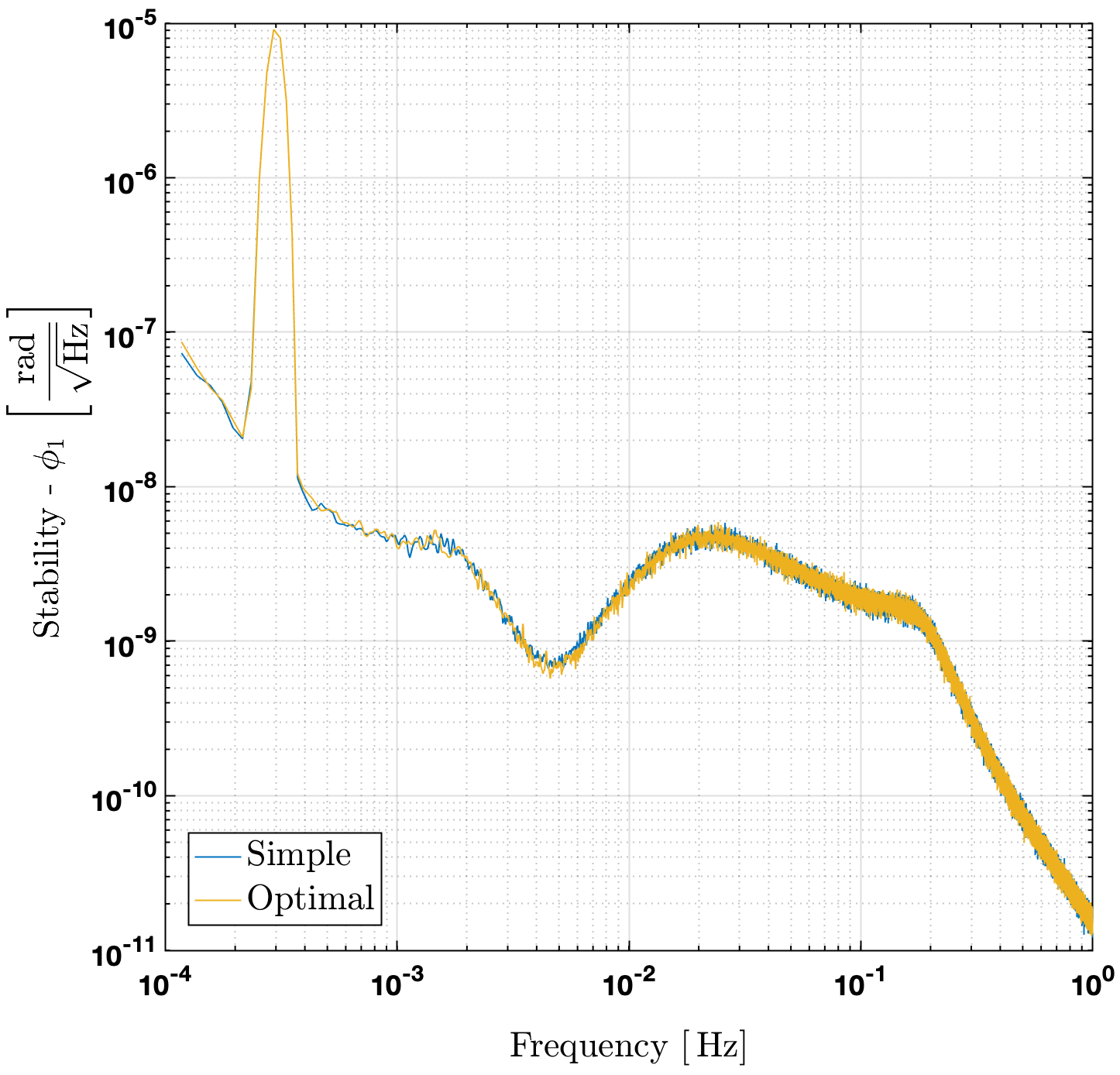}}
    \caption{Stability performance of there representative dynamical d.o.f - $y_1$, $\theta_2$ and $\phi_1$ - using the historical (blue trace) and the optimal (orange trace) suspension schemes. These plots come from the simulation including injection torques on spacecraft at (0.1 \si{\milli\hertz}, 0.2 \si{\milli\hertz} and 0.3 \si{\milli\hertz}). We verify that the stability performance are preserved by the newly proposed suspension scheme.}
\label{figure: AngJitter - Stability}
\end{figure*}

This is visible in plotting the ASD of the test mass positions relative to the spacecraft for $x$, $y$ and $z$ d.o.f.s as shown in Figure\,\myhyperref{figure: AngJitter - Stability}. In each of the sub-figures 
both the case of a simple (blue) and optimized (orange) suspension scheme are shown. We verify that the achieved control performance of both schemes matches very well and therefore we do not see any adverse impact of suspension isolation on the achieved control performance.

\section{Impact of new DFACS scheme on noise budget}
\label{section: budget}

The primary goal of the improved suspension scheme is to mitigate the effect of actuation noise cross-talk into the sensitive $x_1$ and $x_2$ axes due to instrumental imperfections or misalignments. 
An incorrect decoupling of common and differential actuation modes would then in-turn couple this actuation cross-talk to the jitter of the platform, inducing an increased contribution to acceleration noise in the 10-100\,\si{\milli\hertz} frequency band where the drag-free authority is loosest and therefore allows the largest relative motions between the spacecraft and the test masses.

Cross-talk specifications have been set based on LISA Pathfinder experience \cite{bassan_actuation_2018}. Defining $C_{ij}$ as the cross-talk of a force or torque applied along axis $j$ into axis $i$, the values considered are $C_{xy} = C_{xz} = 0.001$, $C_{x\theta} = C_{x\eta} = 0.001\ \si{\metre\per\radian}$ and $C_{x\phi} = 0.005\ \si{\metre\per\radian}$. The asymmetry in angular coefficients is explained by the electrode system geometry, where the electrodes used for $\phi$ actuation 
are located on the $x$ faces of the TM housing. The $C_{x\phi}$ cross-talk element and the contribution from channel $\Phi$ spacecraft jitter to test mass acceleration noise is therefore of particular concern.

\begin{figure}[h!]

\centerline{\includegraphics[scale=0.35, trim={0.0cm 0.0cm 0.0cm 0cm}, clip]{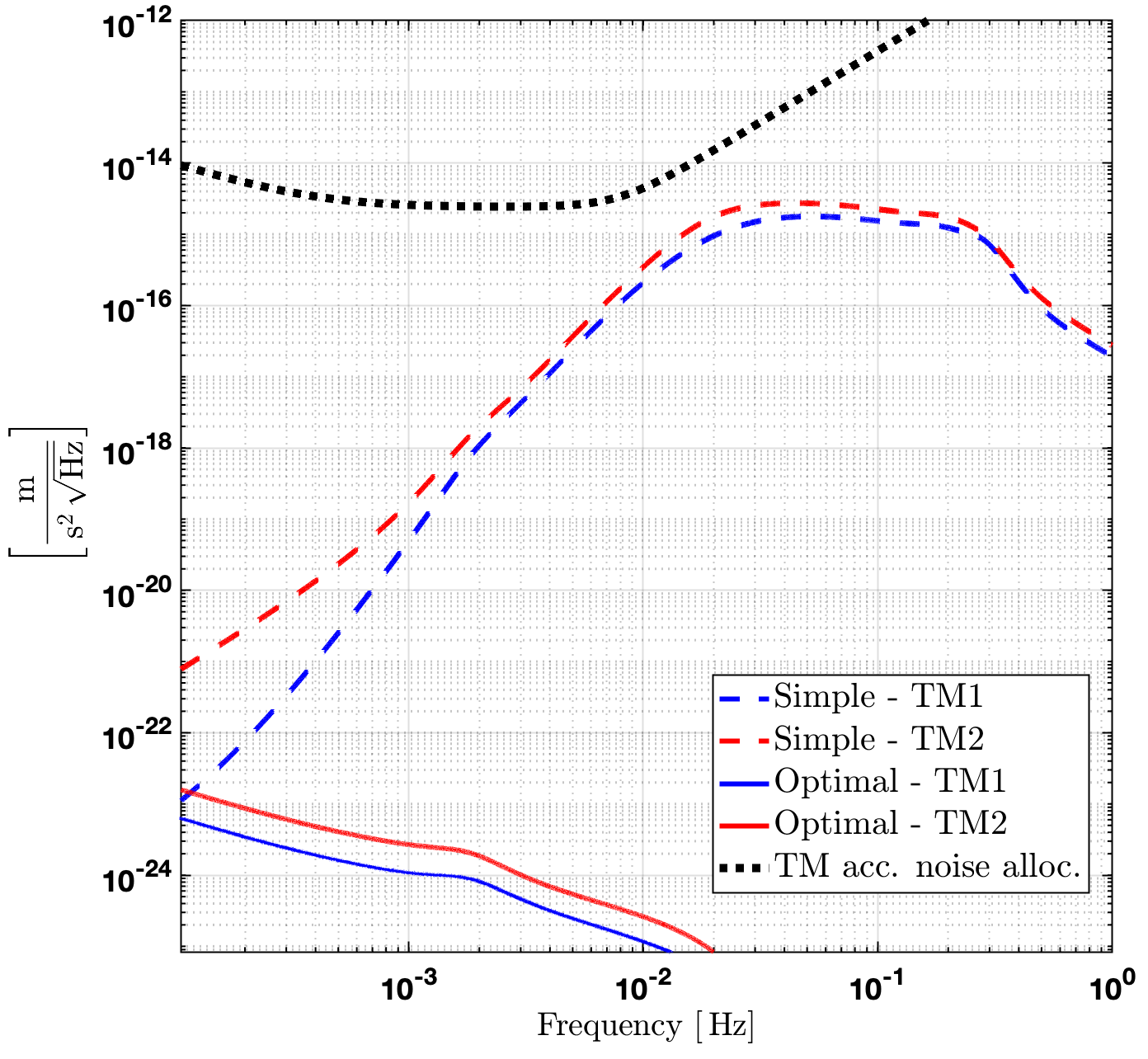}}
\caption{{\it Jitter-induced} actuation cross-talk contributions to acceleration noise budget analyzed with the historical suspensions scheme (dashed lines) and the optimal scheme (solid lines) for the two test masses. The dotted, black lines represent test mass acceleration requirements as a reference.}
\label{figure: ActCont}	

\end{figure}

Fig. \myhyperref{figure: ActCont} shows a comparison between actuation cross-talk contribution to test mass acceleration budget---limiting the analysis to the cross-talk induced by spacecraft jitter---for two suspension schemes considered: the simple scheme (dashed line) using simple one-to-one mapping between dynamical d.o.f. and  force and torque channels, and the optimal scheme (solid lines) developed in this work.
We present results for test mass 1 (in blue) and 2 (in red), together with LISA requirement curve for test mass acceleration (black dotted line). 
We utilize the spacecraft longitudinal and angular jitter requirements as a common-mode disturbance input to the analysis, benefiting from the flexibility of {\tt LISADyn-linear} that allows an easy extraction of specific open-loop transfer functions of interest. Here, the $6 \times 2$ open-loop transfer functions relating the closed-loop spacecraft jitters to the stray forces exerted along $x_1$ and $x_2$ were used to generate the blue and red curves of Fig. \myhyperref{figure: Images/spacecraft}.

Although not dominating any part of the acceleration noise budget, actuation cross-talks can represent a substantial fraction of the noise using the simple suspension scheme. Cross-talk forces driven by spacecraft jitter can reach up to $10\%$ of the budget around the peak of spacecraft jitter amplitude, at 30\,\si{\milli\hertz}, dominated by the $C_{xz}$ and $C_{x\phi}$ coefficients.
Switching to the optimal scheme, one observes in Fig.\,\myhyperref{figure: ActCont} a clear suppression of this effect, leaving only a non-physical residual caused by limitations of numerical precision.
This result is obtained only by rearranging the combinations of sensor data used by the \gls{dfacs} to compute the commanded forces and torques to be applied, that is with no change to sensing or actuation capabilities or control algorithms. 
While the optimization of the suspension will not significantly affect the noise budget, it can yield improved robustness in the DFACS system and suppress cross-talk contributions to a negligible level for minimal cost of implementation.
We also note that this isolation scheme will suppress possible transient suspension force contributions from impulsive events affecting the spacecraft such as micro-meteoroid impacts---numerous such events have been observed and analyzed with LISA Pathfinder \cite{thorpe_micrometeoroid_2019}---since the resulting test-mass to spacecraft motion will be common-mode for both test masses.

\section{Differential mode along the sensitive axes: a local measurement of the differential acceleration noise}
\label{section: acceleration}

At low-frequency, test-mass acceleration noise limits the performance of LISA. Understanding noise behavior is critical to identifying transient events \cite{baghi_detection_2021}, extracting the maximum number of continuous gravitational waveforms and stochastic gravitational wave signals \cite{caprini_reconstructing_2019, flauger_improved_2021}. A number of instrument-noise characterization experiments and calibrations also rely on applying and measuring direct forces on the test masses \cite{armano_calibrating_2018}.
Unlike in other gravitational experiments, there is no way to perfectly ``switch off'' the signal in LISA, the measurement of differential acceleration between test masses in a single LISA arm is dominated by laser frequency noise and TDI combinations are expected to contain continuous gravitational waves with large signal to noise across the LISA frequency band. A local estimate of the differential acceleration noise of the test masses on a single spacecraft immune to both of these effects is therefore valuable.

Following similar logic to that used to derive the optimized suspension scheme already presented, we can derive expressions for the test mass accelerations isolating them from the motion of the spacecraft 
using a combination of suspension forces, GRS and IFO data. 
A development comparable to that of section \myhyperref{subsection: Suspension and differential mode: Longitudinal isolation} allows construction of observables $\Delta a_{x_{1/2}}$ for the differential acceleration between the test masses projected along the drag-free axes---assumed to be the sensitive, long-arm axis to first-order. 
Here we generalize the method of \myhyperref{section: DFACS scheme} and find among all possible combinations, the set of coefficients that minimizes the introduced sensing noise---that which limits the utilization of noisier \gls{grs} channels to the strict minimum required. Such a calculation starts in subtracting from the optical measurement of test mass acceleration along the drag-free axis, $\ddot{x_{1/2}}^\textrm{ifo}$, an estimator of spacecraft acceleration along the same axis, $a_{x_1}^{\text{S/C}}$. The latter can be expressed in terms of a linear combination of the measured motions of the of test masses in their housing reference frames:

\begin{align}
\label{eq: sc motion estimator}
     & a_\indice{x_1}^\exposant{\text{S/C}} = \beta_\indice{11} \ \ddot{y}_1^{\text{grs}} + \alpha_\indice{21} \ \ddot{x}_2^{\text{ifo}} + \beta_\indice{21} \ \ddot{y}_2^{\text{grs}} \\ \nonumber
     & a_\indice{x_2}^\exposant{\text{S/C}} = \beta_\indice{12} \ \ddot{y}_1^{\text{grs}} + \alpha_\indice{12} \ \ddot{x}_1^{\text{ifo}} + \beta_\indice{22} \ y_2^{\text{grs}}, 
\end{align}
where $\alpha$ and $\beta$ are constants to be determined.

Two constraints come directly from geometry: taking as an example test mass $1$, one requires the spacecraft acceleration vector to be co-linear with $\myvec{e}{x_1}$---since one wants to compute $a_\indice{x_1}^\exposant{\text{S/C}} = \vec{a}^\exposant{\ \text{S/C}} \cdot \myvec{e}{x_1}$---and normal to $\myvec{e}{y_1}$, imposing unity norm to the vector combination. The third and last constraint is used to guarantee optimal signal-to-noise ratio. The interferometer $x^{\text{ifo}}$ channels being more than $3$ orders of magnitude more precise than the \gls{grs} channels (see Table \myhyperref{table: NoiseLevels}), the optical noise can be safely neglected here. Hence, the quadratic sum $\beta_\indice{11}^2 S_n^{grs/y_1} + \beta_\indice{21}^2 S_n^{grs/y_2}$ is to be minimized, with $S_n$ denoting the power spectral densities of the two---assumed uncorrelated--- $y^{\text{grs}}$ channels sensing noise. Assuming equal sensing noise levels for $y_1^{\text{grs}}$ and $y_2^{\text{grs}}$ channels, and posing $q = \beta_\indice{11}^2 + \beta_\indice{21}^2$, the optimization translates mathematically as requiring $\tfrac{\text{d} q}{\text{d} \beta_\indice{11}} = 0$, remarking that the geometrical constraints imposes that $\beta_\indice{12} = \tfrac{\sqrt{3} - \beta_\indice{11}}{2}$. With these $3 \times 2$ constraints in total, one can find the following optimal solution:
\begin{align}
\label{eq: coefficients}
     && \alpha_\indice{21} = \frac{4}{5} && \beta_\indice{11} = \frac{\sqrt{3}}{5} && \beta_\indice{21} = \frac{2 \sqrt{3}}{5} \\ \nonumber
     && \alpha_\indice{21} = \frac{4}{5} && \beta_\indice{12} = - \frac{2 \sqrt{3}}{5} && \beta_\indice{22} = -\frac{\sqrt{3}}{5}.
\end{align}

Accounting for lever coupling to spacecraft attitude measured by the long-arm interferometer DWS, subtracting suspension forces applied to the TMs along $y$ to obtain out-of-loop quantities and finally projecting along the axes of interest ($\myvec{e}{x_1}$ for test mass 1, $\myvec{e}{x_2}$ for test mass 2), Equations\,(\myhyperref{eq: diff_acc_est_tm1}) and (\myhyperref{eq: diff_acc_est_tm2}) provide the combinations we finally propose.

\begin{align}
\label{eq: diff_acc_est_tm1}
     \Delta a_{x_\indice{1}}^{\text{est}} & = \; \ddot{x}_1^{\text{ifo}} - \frac{4}{5} \ddot{x}_2^{\text{ifo}} - \frac{\sqrt{3}}{5} \ddot{y}_1^{\text{grs}} - \frac{2 \sqrt{3}}{5} \ddot{y}_2^{\text{grs}} \nonumber \\
     & + \Delta z \ddot{\Theta}^\sensor{ldws} - \frac{\Delta z}{2} \ddot{H}^\sensor{ldws} - \left( \Delta x - \frac{\Delta y}{\sqrt{3}} \right) \ddot{\Phi}^\sensor{ldws} \nonumber \\
     & + \frac{\sqrt{3}}{5} \frac{F_{y_\indice{1}}^{\text{sus}}}{m_1} + \frac{2 \sqrt{3}}{5} \frac{F_{y_\indice{2}}^{\text{sus}}}{m_2}
     \\
\label{eq: diff_acc_est_tm2}
     \Delta a_{x_\indice{2}}^{\text{est}} & = \; \ddot{x}_2^{\text{ifo}} - \frac{4}{5} \ddot{x}_2^{\text{ifo}} + \frac{2 \sqrt{3}}{5} \ddot{y}_1^{\text{grs}} + \frac{\sqrt{3}}{5} \ddot{y}_2^{\text{grs}} \nonumber \\
     & + \Delta z \ddot{\Theta}^\sensor{ldws} + \frac{\Delta z}{2} \ddot{H}^\sensor{ldws} - \left( \Delta x + \frac{\Delta y}{\sqrt{3}} \right) \ddot{\Phi}^\sensor{ldws} \nonumber \\
     & - \frac{2 \sqrt{3}}{5} \frac{F_{y_\indice{1}}^{\text{sus}}}{m_1} - \frac{\sqrt{3}}{5} \frac{F_{y_\indice{2}}^{\text{sus}}}{m_2}
\end{align}

Note that in combining \gls{ifo}, \gls{grs} and \gls{ldws} information, the estimate precision is deteriorated relative to the true test mass motion by \gls{ldws} and \gls{grs} sensing noise. At low frequency however, where the sensing noise have a lesser impact on acceleration noise, the estimator becomes very accurate. 

\begin{figure*}[htb]
    \centering
    \subfloat[Time series: comparison between estimator and true acceleration noise for an injected signal with an amplitude of $0.5\times 10^{-12}$\,m\,s$^{-2}$ at 0.1\,mHz. Both quantities have been processed identically through low-pass filtering (order 3, $f_c = 2 \si{\milli\hertz}$) and detrending in order to highlight their matching. \label{figure: EstAcc/TimeseriesA}]
    {\includegraphics[scale=0.35, trim={0.0cm 0.0cm 0.0cm 0.0cm}, clip]{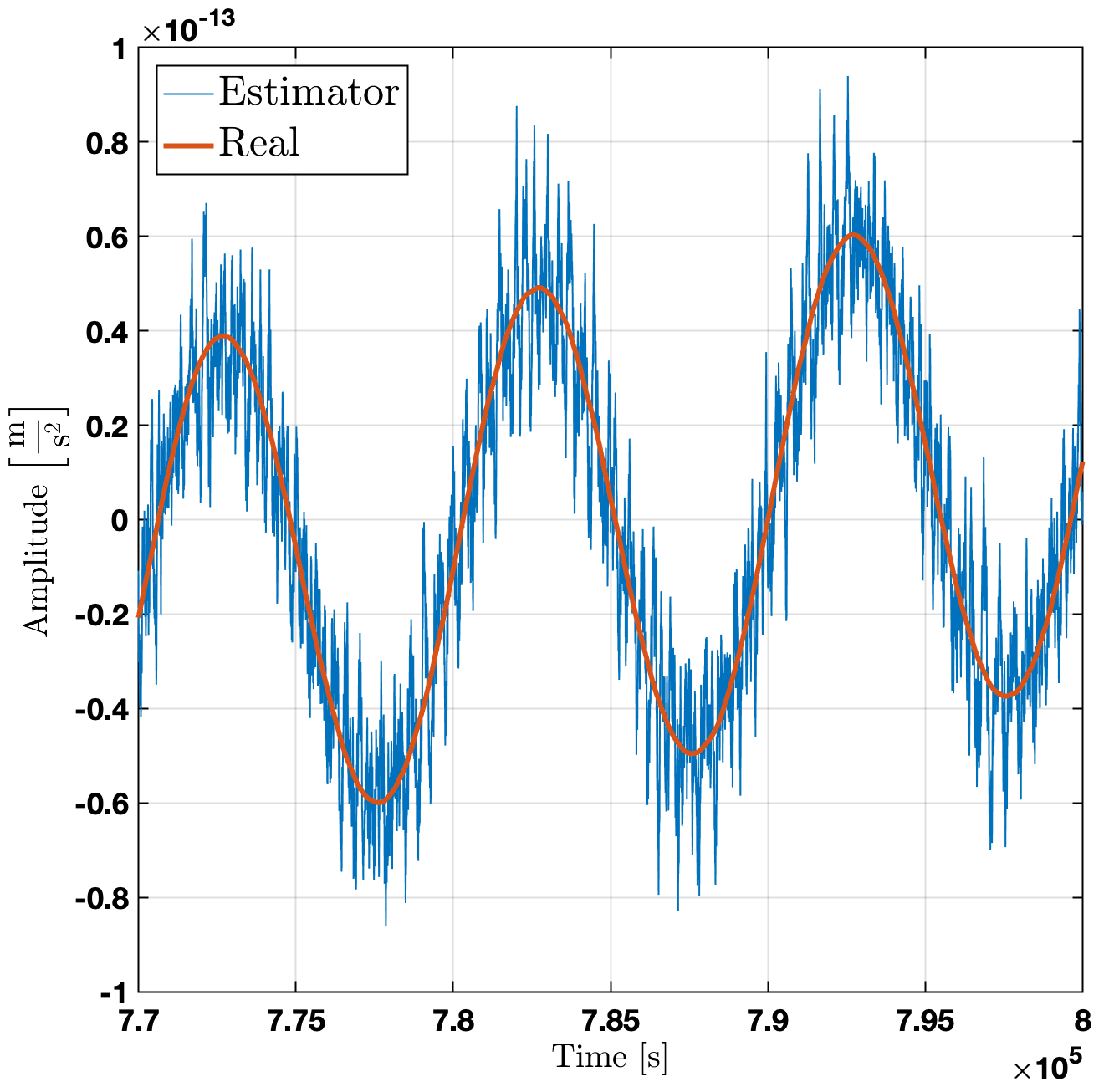}}
    \hspace{0.3cm}
    \subfloat[Spectrum breakdown of the estimator and comparison to true acceleration. In blue the total estimator, in red the true quantity to be measured, in yellow the residual, in purple the contribution from the sensor channels combination and in green from the commanded forces along $y_1$ and $y_2$ \label{figure: EstAcc/TimeseriesB}. Light blue trace is the true differential acceleration between the two test masses projected along $\myvec{e}{x_1}$.]
    {\includegraphics[scale=0.35, trim={0.0cm 0.0cm 0.0cm 0.0cm}, clip]{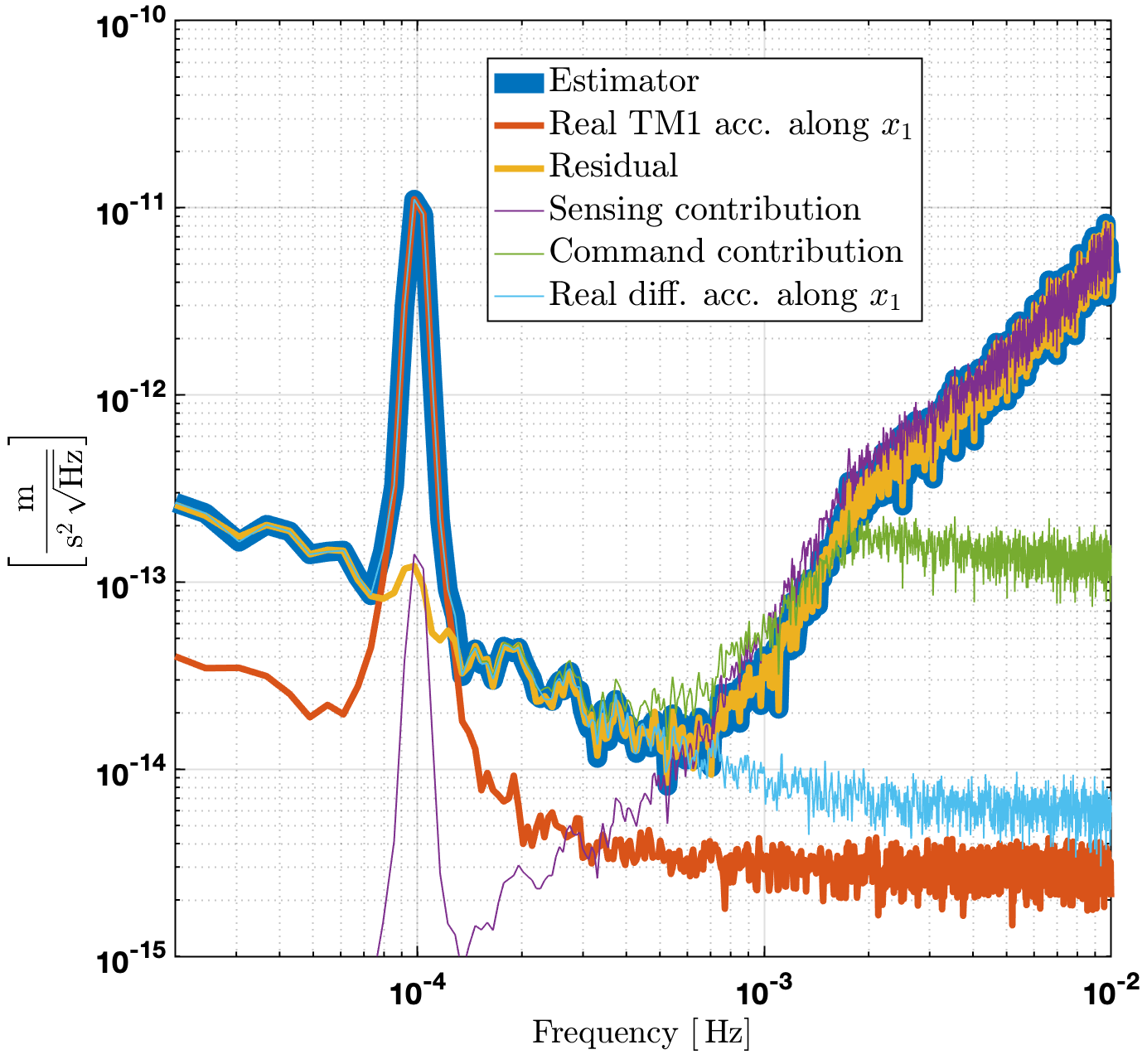}}
    \caption{Estimation of the differential acceleration between test masses projected along $\myvec{e}{x_1}$ from on-board sensors (\gls{ifo}, \gls{grs} and \gls{ldws}) and commanded forces.}
\label{figure: acc_est}
\end{figure*}

We present in Fig.\,\myhyperref{figure: acc_est} the result of a simulation experiment intended to compare the acceleration estimator written in Equation\,(\myhyperref{eq: diff_acc_est_tm1}) to the true acceleration of test masses $1$ and $2$ projected along their respective axes---physical quantities obviously not available to observers but known by the simulators and its users. A sinusoidal force of $0.1\,\si{\pico\newton}$ is injected on test mass 1 at the $f=0.1\,\si{\milli\hertz}$. Time series traces on the left-hand sub-figure show that the estimator is able to recover the injected signal well, while the spectra on the right-hand side shows that there is no significant residual left over after subtraction. The spectrum breakdown illustrates the performance of the estimator. Below 0.6\,mHz the estimator noise is a factor 5-6 above the target acceleration noise level. The additional noise originates from the the acceleration noise of the second test mass along its $y$ axis, projected onto the first TM $x$ axis hence introducing significant additional noise. In fact, the signal resolved by the estimator below $0.6\,\si{\milli\hertz}$ is exactly the {\it differential acceleration} between the two test masses along $\myvec{e}{x_1}$---and respectively $\myvec{e}{x_2}$)---drawn in light blue in Fig.\,\myhyperref{figure: acc_est}. Expressions for these quantities are given in Equations \,(\myhyperref{eq: diff_acc_true_tm1}) and (\myhyperref{eq: diff_acc_true_tm2}). Above $0.6 \si{\milli\hertz}$, \gls{grs}, and to a lesser extent, \gls{ldws} sensing noise take over as leading contributors to noise.

\begin{align}
\label{eq: diff_acc_true_tm1}
     \Delta a_{x_\indice{1}}^{\text{true}} = & \; a_{x_\indice{1}}^{\text{true}} - \frac{4}{5} a_{x_\indice{2}}^{\text{true}} - \frac{\sqrt{3}}{5} \left( a_{y_\indice{1}}^{\text{true}} - \frac{F_{y_\indice{1}}^{\text{sus}}}{m_1} \right)
     \nonumber \\
     & - \frac{2 \sqrt{3}}{5} \left( a_{y_\indice{2}}^{\text{true}} - \frac{F_{y_\indice{2}}^{\text{sus}}}{m_2} \right)
     \\
\label{eq: diff_acc_true_tm2}
     \Delta a_{x_\indice{2}}^{\text{true}} = & \; a_{x_\indice{2}}^{\text{true}} - \frac{4}{5} a_{x_\indice{1}}^{\text{true}} + \frac{2 \sqrt{3}}{5} \left( a_{y_\indice{1}}^{\text{true}} - \frac{F_{y_\indice{1}}^{\text{sus}}}{m_1} \right)
     \nonumber \\
     & + \frac{\sqrt{3}}{5} \left( a_{y_\indice{2}}^{\text{true}} - \frac{F_{y_\indice{2}}^{\text{sus}}}{m_2} \right) 
\end{align}

\section{Conclusion}

We have proposed optimized suspension control coordinates for the LISA Drag Free and Attitude Control System in order to isolate commanded forces and torques applied on test masses from spacecraft jitter. 
Using these coordinates, we have demonstrated the isolation efficiency by simulating the excitation of the spacecraft longitudinally and rotationally and observing its rejection in the response on the suspension command. Meanwhile, we verify that the new suspension coordinates do not deteriorate stability performance while keeping test masses locked at their set points in response to the attitude guidance produced by the the annual rotation of LISA constellation. An important consequence of the new suspension scheme is a significant reduction of the level of suspension forces and torques applied to the test masses, and subsequently a large mitigation of actuation cross-talk effect which would lead to important contribution to the acceleration noise budget. On the other hand, coupling through dynamical stiffness or tilt-to-length stay unchanged since stability performance are similar relative to the former, simple scheme usually considered.

A possible disadvantage of the new scheme may be an increase of the complexity of the control scheme and the of the sensor mapping, making in-loop sensor outputs and calibration experiments interpretation less immediate. Table \myhyperref{table: Coupling Matrix} provides all the information needed about such mapping, and from it, basic, analytical processing is needed to rotate coordinates back to more understandable quantities. In addition, this new control scheme relies extensively on multi-sensor fusion  and this can have robustness implications regarding sensor failures. However, sensor failure may have critical impact on any control scheme, including the simpler scheme of Appendix \myhyperref{appendix: simple scheme}. Further analyses will be needed in order to understand impact of failure for the various control schemes considered. It is important to note that such novel scheme proposition involves software implementation only, and addresses the particular case of a science mode in nominal operational conditions. It will always be possible to modify the control scheme on-board the spacecraft in case of failure or for any operational reason. It is to be expected that various control modes will be developed and available in flight for LISA, for the different phases of the missions or in case of hardware failure.

Based on similar reasoning and algebra, we have proposed estimators of the differential acceleration (projected along interferometer arms) between the two local test-masses on-board LISA spacecraft. We verify by simulation that this estimate agrees with true acceleration quantities below 0.6\,\si{\milli\hertz}, while being limited above that frequency due to \gls{grs} sensing noise. This composite data product provides critical information about local test mass accelerations which will have important use for data sanity and quality check, calibration and data processing, data analysis and data artifact corrections.

\appendix

\section{Note on the simple DFACS scheme}
\label{appendix: simple scheme}

The simple control scheme, with direct suspension on \gls{grs} channels, captures necessarily the noisy motion of the spacecraft. This simple strategy then suffers from a coupling between the suspension command and spacecraft longitudinal and rotational jitter. As in section \myhyperref{section: DFACS scheme} for the optimized scheme, here are detailed the $15$ control coordinates $[ \hat{x}_\indice{1}^\exposant{\text{sim}}, \allowbreak  \hat{x}_\indice{2}^\exposant{\text{sim}}, \allowbreak  \hat{z}_\indice{1}^\exposant{\text{sim}}, \allowbreak  \hat{\Theta}^\exposant{\text{sim}}, \allowbreak  \hat{H}^\exposant{\text{sim}}, \allowbreak  \hat{\Phi}^\exposant{\text{sim}}, \allowbreak  \hat{y}_\indice{1}^\exposant{\text{sim}}, \allowbreak  \hat{y}_\indice{2}^\exposant{\text{sim}}, \allowbreak  \hat{z}_\indice{2}^\exposant{\text{sim}}, \allowbreak  \hat{\theta}_\indice{1}^\exposant{\text{sim}}, \allowbreak  \hat{\eta}_\indice{1}^\exposant{\text{sim}}, \allowbreak  \hat{\phi}_\indice{1}^\exposant{\text{sim}}, \allowbreak  \hat{\theta}_\indice{2}^\exposant{\text{sim}}, \allowbreak  \hat{\eta}_\indice{2}^\exposant{\text{sim}}, \allowbreak  \hat{\phi}_\indice{2}^\exposant{\text{sim}}]$ which defines the simple \gls{dfacs} scheme we refer to throughout the paper and formerly considered on initial technical work and internal report on LISA \gls{dfacs} by industry. Equation \myhyperref{eq: DragFreeScheme - Simple} and \myhyperref{eq: AttitudeScheme - Simple} gives drag-free and attitude control coordinates while \myhyperref{eq: scheme_sus_long - Simple} and \myhyperref{eq: scheme_sus_angle_tm - Simple} provides both suspension longitudinal and angular coordinates. Table \myhyperref{table: simple scheme} gives an overall, matrix view of the control strategy.

\begin{align}
\label{eq: DragFreeScheme - Simple}
    & F_{X}^{\text{drag-free}} \propto \hat{x}_\indice{1}^\exposant{\text{sim}} \equiv x_1^\sensor{ifo} \nonumber \\
    & F_{Y}^{\text{drag-free}} \propto \hat{x}_\indice{2}^\exposant{\text{sim}} \equiv x_2^\sensor{ifo} \\
    & F_{Z}^{\text{drag-free}} \propto \hat{z}_\indice{1}^\exposant{\text{sim}} \equiv z_1^\sensor{grs} \nonumber
\end{align}

\begin{align}
\label{eq: AttitudeScheme - Simple}
    & N_{X}^{\text{att}} \propto \hat{\Theta}^\exposant{\text{sim}} \equiv \Theta^\sensor{ldws} = \eta_2^\sensor{ldws} - \eta_1^\sensor{ldws} \nonumber \\
    & N_{Y}^{\text{att}} \propto \hat{H}^\exposant{\text{sim}} \equiv H^\sensor{ldws} = -\frac{1}{\sqrt{3}} \left( \eta_1^\sensor{ldws} + \eta_2^\sensor{ldws} \right) \\
    & N_{Z}^{\text{att}} \propto \hat{\Phi}^\exposant{\text{sim}} \equiv \Phi^\sensor{ldws} = -\frac{1}{2} \left( \phi_1^\sensor{ldws} + \phi_2^\sensor{ldws} \right) \nonumber
\end{align}

\begin{align}
\label{eq: scheme_sus_long - Simple}
     & F_{y_1}^{\text{sus}} \propto
     \hat{y}_\indice{1}^\exposant{\text{sim}} \equiv y_1^\sensor{grs} \nonumber \\[0.2in]
     & F_{y_2}^{\text{sus}} \propto \hat{y}_\indice{2}^\exposant{\text{sim}} \equiv y_2^\sensor{grs} \\[0.2in]
     & F_{z_1}^{\text{sus}} \propto \hat{z}_\indice{1}^\exposant{\text{sim}} \equiv 0 \\[0.2in]
     & F_{z_2}^{\text{sus}} \propto \hat{z}_\indice{2}^\exposant{\text{sim}} \equiv z_2^\sensor{grs} \nonumber
\end{align}

\begin{align}
\label{eq: scheme_sus_angle_tm - Simple}
    & N_{x_1}^{\text{sus}} \propto \hat{\theta}_\indice{1}^\exposant{\text{sim}} \equiv \theta_1^\sensor{grs}
    \nonumber \\
    & N_{y_1}^{\text{sus}} \propto \hat{\eta}_\indice{1}^\exposant{\text{sim}} \equiv \eta_1^\sensor{ifo}
    \nonumber \\
    & N_{z_1}^{\text{sus}} \propto \hat{\phi}_\indice{1}^\exposant{\text{sim}} \equiv \phi_1^\sensor{ifo}
\\[0.2in]
    & N_{x_2}^{\text{sus}} \propto \hat{\theta}_\indice{2}^\exposant{\text{sim}} \equiv \theta_2^\sensor{grs}
    \nonumber \\
    & N_{y_2}^{\text{sus}} \propto \hat{\eta}_\indice{2}^\exposant{\text{sim}} \equiv \eta_2^\sensor{ifo}
    \nonumber \\
    & N_{z_2}^{\text{sus}} \propto \hat{\phi}_\indice{2}^\exposant{\text{sim}} \equiv \phi_2^\sensor{ifo}
    \nonumber
\end{align}

\begin{table*}[t]
\centering
\caption{DFACS simple scheme represented through the mapping matrix from the sensing error signals to the commanded forces / torques. The symbol "-" replaces the $0.0$ value to enhance readability of the table.}
\label{table: simple scheme}
\begin{tabular}{cc||c|c|c|c|c|c|c|c|c|c|c|c|c|c|c|c|c|c|c|c|c|}
\multicolumn{2}{c||}{Control} & $\Theta^\sensor{ldws}$ & $H^\sensor{ldws}$ & $\Phi^\sensor{ldws}$ & $x_\indice{1}^\sensor{ifo}$ & $\eta_\indice{1}^\sensor{ifo}$ & $\phi_\indice{1}^\sensor{ifo}$ & $x_\indice{2}^\sensor{ifo}$ & $\eta_\indice{2}^\sensor{ifo}$ & $\phi_\indice{2}^\sensor{ifo}$ & $x_\indice{1}^\sensor{grs}$ & $y_\indice{1}^\sensor{grs}$ & $z_\indice{1}^\sensor{grs}$ & $\theta_\indice{1}^\sensor{grs}$ & $\eta_\indice{1}^\sensor{grs}$ & $\phi_\indice{1}^\sensor{grs}$ & $x_\indice{2}^\sensor{grs}$ & $y_\indice{2}^\sensor{grs}$ & $z_\indice{2}^\sensor{grs}$ & $\theta_\indice{2}^\sensor{grs}$ & $\eta_\indice{2}^\sensor{grs}$ & $\phi_\indice{2}^\sensor{grs}$ \\
\hline
\noalign{\vskip 2mm}
\hline
\multirow{3}*{Att.}
    & $N_X$ & $1$ & {-} & {-} & {-} & {-} & {-} & {-} & {-} & {-} & {-} & {-} & {-} & {-} & {-} & {-} & {-} & {-} & {-} & {-} & {-} & {-} \\
    \cline{2-23}
    & $N_Y$ & {-} & $1$ & {-} & {-} & {-} & {-} & {-} & {-} & {-} & {-} & {-} & {-} & {-} & {-} & {-} & {-} & {-} & {-} & {-} & {-} & {-} \\
    \cline{2-23}
    & $N_Z$ & {-} & {-} & $1$ & {-} & {-} & {-} & {-} & {-} & {-} & {-} & {-} & {-} & {-} & {-} & {-} & {-} & {-} & {-} & {-} & {-} & {-} \\
\hline
\noalign{\vskip 8mm}
\hline
\multirow{3}*{DF}
    & $F_X$ & {-} & {-} & {-} & $1$ & {-} & {-} & {-} & {-} & {-} & {-} & {-} & {-} & {-} & {-} & {-} & {-} & {-} & {-} & {-} & {-} & {-} \\
    \cline{2-23}
    & $F_Y$ & {-} & {-} & {-} & {-} & {-} & {-} & $1$ & {-} & {-} & {-} & {-} & {-} & {-} & {-} & {-} & {-} & {-} & {-} & {-} & {-} & {-} \\
    \cline{2-23}
    & $F_Z$ & {-} & {-} & {-} & {-} & {-} & {-} & {-} & {-} & {-} & {-} & {-} & $1$ & {-} & {-} & {-} & {-} & {-} & {-} & {-} & {-} & {-} \\
\hline
\noalign{\vskip 8mm}
\hline
    & $F_{x_\indice{1}}$ & {-} & {-} & {-} & {-} & {-} & {-} & {-} & {-} & {-} & {-} & {-} & {-} & {-} & {-} & {-} & {-} & {-} & {-} & {-} & {-} & {-} \\
    \cline{2-23}
\multirow{3}*{Sus.}
    & $F_{y_\indice{1}}$ & {-} & {-} & {-} & {-} & {-} & {-} & {-} & {-} & {-} & {-} & $1$ & {-} & {-} & {-} & {-} & {-} & {-} & {-} & {-} & {-} & {-} \\
    \cline{2-23}
\multirow{3}*{TM1}
    & $F_{z_\indice{1}}$ & {-} & {-} & {-} & {-} & {-} & {-} & {-} & {-} & {-} & {-} & {-} & {-} & {-} & {-} & {-} & {-} & {-} & {-} & {-} & {-} & {-} \\
    \cline{2-23}
    & $N_{x_\indice{1}}$ & {-} & {-} & {-} & {-} & {-} & {-} & {-} & {-} & {-} & {-} & {-} & {-} & $1$ & {-} & {-} & {-} & {-} & {-} & {-} & {-} & {-} \\
    \cline{2-23}
    & $N_{y_\indice{1}}$ & {-} & {-} & {-} & {-} & {-} & {-} & {-} & {-} & {-} & {-} & {-} & {-} & {-} & $1$ & {-} & {-} & {-} & {-} & {-} & {-} & {-} \\
    \cline{2-23}
    & $N_{z_\indice{1}}$ & {-} & {-} & {-} & {-} & {-} & {-} & {-} & {-} & {-} & {-} & {-} & {-} & {-} & {-} & $1$ & {-} & {-} & {-} & {-} & {-} & {-} \\
\hline
\noalign{\vskip 4mm}
\hline
    & $F_{x_\indice{2}}$ & {-} & {-} & {-} & {-} & {-} & {-} & {-} & {-} & {-} & {-} & {-} & {-} & {-} & {-} & {-} & {-} & {-} & {-} & {-} & {-} & {-} \\
    \cline{2-23}
\multirow{3}*{Sus.}
    & $F_{y_\indice{2}}$ & {-} & {-} & {-} & {-} & {-} & {-} & {-} & {-} & {-} & {-} & {-} & {-} & {-} & {-} & {-} & {-} & $1$ & {-} & {-} & {-} & {-} \\
    \cline{2-23}
\multirow{3}*{TM2}
    & $F_{z_\indice{2}}$ & {-} & {-} & {-} & {-} & {-} & {-} & {-} & {-} & {-} & {-} & {-} & {-} & {-} & {-} & {-} & {-} & {-} & $1$ & {-} & {-} & {-} \\
    \cline{2-23}
    & $N_{x_\indice{2}}$ & {-} & {-} & {-} & {-} & {-} & {-} & {-} & {-} & {-} & {-} & {-} & {-} & {-} & {-} & {-} & {-} & {-} & {-} & $1$ & {-} & {-} \\
    \cline{2-23}
    & $N_{y_\indice{2}}$ & {-} & {-} & {-} & {-} & {-} & {-} & {-} & {-} & {-} & {-} & {-} & {-} & {-} & {-} & {-} & {-} & {-} & {-} & {-} & $1$ & {-} \\
    \cline{2-23}
    & $N_{z_\indice{2}}$ & {-} & {-} & {-} & {-} & {-} & {-} & {-} & {-} & {-} & {-} & {-} & {-} & {-} & {-} & {-} & {-} & {-} & {-} & {-} & {-} & $1$ \\
\hline
\end{tabular}
\end{table*}

\section{Acknowledgement}
\label{S:ack}

Henri Inchausp\'e would like to acknowledge the Centre Nationale d'\'Etudes Spatiales (CNES) for its financial support. Peter Wass, Orion Sauter and Henri Inchausp\'e were supported by NASA LISA Preparatory Science program, grant number 80NSSC19K0324.

% \printglossary[type=\acronymtype]
\bibliographystyle{unsrt2}
\bibliography{references.bib}
 \end{document}